\documentclass[10pt,twocolumn,letterpaper]{article}

\usepackage{iccv}
\usepackage{times}
\usepackage{epsfig}
\usepackage{graphicx}
\usepackage{amsmath}
\usepackage{amssymb}
\usepackage{subcaption,siunitx,booktabs}
\usepackage[accsupp]{axessibility}

\usepackage{framed,multirow}

\usepackage{amssymb,amssymb,amsfonts}
\usepackage{latexsym}
\usepackage{algorithmic}
\usepackage{graphicx}
\usepackage{textcomp}
\usepackage{caption}
\usepackage{multirow}
\usepackage{float}
\usepackage{breqn}
\usepackage{subcaption}
\usepackage{xcolor}
\usepackage{adjustbox}
\usepackage{array}
\usepackage{multirow}
\usepackage{makecell}
\usepackage{times}
\usepackage{epsfig}
\usepackage{amsmath}
\usepackage{amssymb}
\usepackage{subcaption,siunitx,booktabs}
\usepackage{rotating}
\usepackage{array,multirow}
\usepackage{arydshln}
\usepackage{soul}

\usepackage{pifont}

\newcommand\blfootnote[1]{%
  \begingroup
  \renewcommand\thefootnote{}\footnote{#1}%
  \addtocounter{footnote}{-1}%
  \endgroup
}

\usepackage[pagebackref=true,breaklinks=true,letterpaper=true,colorlinks,bookmarks=false]{hyperref}

\iccvfinalcopy 

\def\iccvPaperID{32} 

\ificcvfinal\fi

\begin{document}

\title{One Model is All You Need: Multi-Task Learning Enables Simultaneous Histology Image Segmentation and Classification}

\author{Simon Graham$^{1,3}$, Quoc Dang Vu$^1$, Mostafa Jahanifar$^1$, Shan E Ahmed Raza$^1$, \\ Fayyaz Minhas$^1$, David Snead$^{2,3}$ and Nasir Rajpoot$^{1,2,3}$ \\ \\
$^1$Department of Computer Science, University of Warwick, UK \\
$^2$Department of Pathology, University Hospitals Coventry and Warwickshire NHS Trust, UK \\
$^3$Histofy Ltd, UK \\
{\tt\small simon.graham@warwick.ac.uk}
}

\maketitle
\thispagestyle{empty}

\begin{abstract}
The recent surge in performance for image analysis of digitised pathology slides can largely be attributed to the advances in deep learning. Deep models can be used to initially localise various structures in the tissue and hence facilitate the extraction of interpretable features for biomarker discovery. However, these models are typically trained for a single task and therefore scale poorly as we wish to adapt the model for an increasing number of different tasks. Also, supervised deep learning models are very \textit{data hungry} and therefore rely on large amounts of training data to perform well. In this paper, we present a multi-task learning approach for segmentation and classification of nuclei, glands, lumina and different tissue regions that leverages data from multiple independent data sources. While ensuring that our tasks are aligned by the same tissue type and resolution, we enable meaningful simultaneous prediction with a single network. As a result of feature sharing, we also show that the learned representation can be used to improve the performance of additional tasks via transfer learning, including nuclear classification and signet ring cell detection. As part of this work, we train our developed \textit{Cerberus} model on a huge amount of data, consisting of over 600K objects for segmentation and 440K patches for classification. We use our approach to process 599 colorectal whole-slide images from TCGA, where we localise 377 million, 900K and 2.1 million nuclei, glands and lumina, respectively and make the results available to the community for downstream analysis.
\end{abstract}


\section{Introduction}

In recent years, there has been a progressive shift towards the digitisation of histopathology slides, enabling the development of computer vision (CV) techniques for automated tissue analysis. In particular, with the rapidly increasing amount of pathology image data and computing power, deep learning has revolutionised the field of computational pathology (CPath). A large proportion of deep CV models used in CPath are convolutional neural networks (CNNs) and have proven to be successful when applied to a plethora of tasks, including cancer detection \cite{bejnordi2017diagnostic, veeling2018rotation}, cancer grading \cite{bulten2020automated, shaban2020context} and survival analysis \cite{bychkov2018deep, shaban2019novel}. However, \textit{black box} CNNs may have a poor level of interpretability when used directly in slide classification, as they are typically non-linear and are governed by millions of parameters. 

To help overcome the above challenge, deep learning can be leveraged as an initial step to localise regions of interest in the tissue before extracting human-interpretable and biologically meaningful features for downstream slide analysis. This fosters the development of more explainable pipelines for CPath, which is particularly important for boosting the confidence of pathologists in their diagnoses. Explainable models increase model transparency by providing an output that is easy to understand and therefore may be utilised as a tool to help facilitate the decision of the pathologist. Furthermore, using human-interpretable features \cite{diao2021human} in explainable pipelines can help reveal new biomarkers for complex tasks, such as survival analysis and image-based detection of clinically actionable genetic alterations \cite{kather2020pan}. 

Previous works on the localisation of various regions and constructs in the tissue typically involve the use of a devoted CNN for a particular task. For example, Diao \textit{et al.}  \cite{diao2021human} trained a network to detect and classify nuclei and an additional network for the segmentation of different tissue types before feature extraction. This strategy is often needed because conventional training pipelines require target objects to be labelled across the entire dataset. Optimising networks separately also ensures that training procedures do not need to be tuned over multiple tasks. However, using one model per task may not be scalable as we wish to localise an increasing number of tissue constructs. Instead, utilising a single model that can perform multiple tasks \textit{simultaneously} is preferable because it reduces the computational overhead, can benefit from shared features between tasks and also saves the time needed for training separate models. 

Multi-Task Learning (MTL) \cite{caruana1997multitask, zhang2017survey} has previously been explored in CPath, but with a key focus on obtaining a generalisable encoder \cite{gamper2020multi} for transfer learning \cite{mormont2020multi} or image compression \cite{tellez2020extending}. Here, each task may include image regions from different tissue types, with varying resolution and potentially with different staining. Therefore, such models can not be used to perform simultaneous prediction for a single input image, due to inconsistency between assumed inputs for each task. Instead, it may be beneficial to utilise an MTL approach with \textit{aligned} tasks, thereby enabling simultaneous prediction with a single model. Not only does this enable a meaningful simultaneous prediction, but using aligned tasks also helps ease optimisation by reducing the potential for conflicting gradients to exist between tasks. Therefore, this holds potential for achieving a strong performance across all considered tasks, which was previously not possible with other MTL approaches in CPath \cite{gamper2020multi}.

To perform well, deep learning approaches require an abundance of labelled data, which is a current bottleneck in the development of CPath models because the collection of annotations requires significant pathologist input \cite{wahab2021semantic, tizhoosh2018artificial}. Without large and diverse datasets, any developed model may struggle to generalise to unseen examples, limiting potential usage in a clinical setting. MTL may help to overcome this challenge by utilising a shared encoder, enabling features to be learned over multiple tasks \cite{crawshaw2020multi}. Yet, it is desirable for each task to have access to a large amount of annotated data from multiple sources to ensure strong performance.

In this paper, we propose a multi-task learning approach for simultaneous segmentation and classification of nuclei, glands, lumina and different tissue regions. With the use of a novel sampling strategy, the proposed \textit{Cerberus} model leverages data from multiple independent sources during training, enabling a competitive performance compared to single-task alternatives while only requiring a single network. Therefore, our approach is the first MTL method in CPath that can sustain a strong performance at the output of the network across all considered tasks. We demonstrate that Cerberus learns a strong feature representation that can help improve the performance of additional tasks via transfer learning, including nuclear classification and signet ring cell detection. To ensure that Cerberus has sufficient labelled examples from multiple different data sources, we create a large and diverse colorectal dataset containing over 600 thousand object boundaries for segmentation and 440 thousand patches for classification. In addition, we process all colorectal slides from The Cancer Genome Atlas (TCGA) and make the results accessible. We hope that doing this will remove a major barrier for the development of explainable approaches for CPath.

The main contributions of this work are listed as follows\blfootnote{Cerberus model code and results on the TCGA dataset can be found here: \url{https://github.com/TissueImageAnalytics/cerberus}}:

\begin{itemize}
    \item We present a multi-task network, named Cerberus, that is capable of performing simultaneous prediction with comparable performance to single-task learning methods.
    \item We show that Cerberus learns a strong feature representation that can be leveraged during transfer learning to boost the performance of additional tasks.
    \item We use our method to process all colorectal slides from TCGA, resulting in 377 million, 900 thousand and 2.1 million nuclei, glands and lumina respectively and make the results available.
\end{itemize}

\section{Related Work}
\subsection{Multi-task learning vs multi-label learning}
\label{section:mtl_vs_mll}
 MTL for computer vision enables multiple independent datasets to be leveraged during training time, which offers advantages like improved data efficiency, reduced overfitting through shared representations and fast learning by leveraging auxiliary information \cite{crawshaw2020multi}. There are various ways in which MTL can be implemented, including weight sharing \cite{caruana1997multitask, mormont2020multi} and architecture adaptations \cite{strezoski2019many, misra2016cross}. For deep learning, MTL is typically done via hard or soft parameter sharing of hidden layers \cite{ruder2017overview}. Hard parameter sharing is the most common strategy used for MTL, where hidden layers are shared between tasks before using several task-specific output layers. In soft parameter sharing, each task has its own model, where the associated parameters are constrained \cite{duong2015low, yang2016trace} so that they are similar for each task. Therefore, using hard parameter sharing may be preferable because using separate models may not be feasible when the number of tasks is large. 
 
\begin{figure}[t]
		\centering
        \includegraphics[width=1.0\columnwidth]{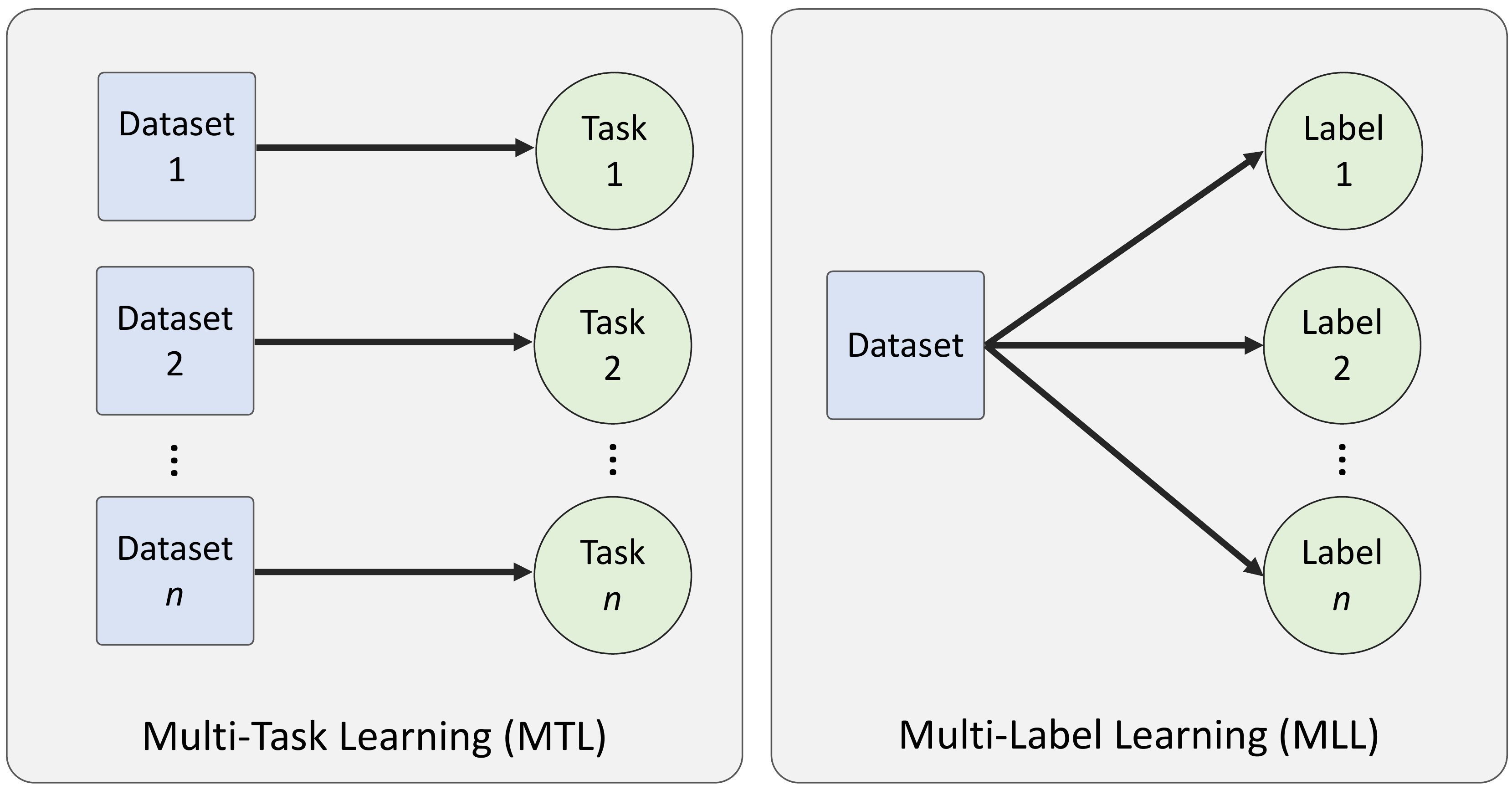}
		\caption{Multi-Task Learning (MTL) vs Multi-Label Learning (MLL): MTL is a training paradigm where a model is trained over different predictive tasks. Here, labels are task-dependent. MLL requires labels to be provided for every input sample.}
		\label{fig:mtl_vs_mll}
	\end{figure}

 It is important to distinguish MTL from Multi-Label Learning (MLL) \cite{zhang2013review}, where each data point is associated with multiple labels. If each label is treated as a separate task, MLL can be viewed as a special case of MTL where different tasks always share the same data \cite{zhang2017survey}. For clarity, we show the differences between these two learning strategies in Figure \ref{fig:mtl_vs_mll}. 
 
\subsection{Multi-label learning for CPath}
As mentioned above, it is possible to predict multiple tasks at the output of the network using MLL. For example, Shephard \textit{et al.} \cite{shephard2021simultaneous} performed nuclear segmentation and classification along with segmentation of intra-epithelial layers in oral tissue. Similarly, Fraz \textit{et al.} \cite{fraz2020fabnet} performed simultaneous segmentation of blood vessels and nerves. However, because the two above-mentioned approaches use MLL, they expect every input image to have associated labels for all tasks. This limits the data that can be used because it is common for available datasets in CPath to only consider a single object \cite{sirinukunwattana2017gland, gamper2020pannuke} or label \cite{kather2019predicting}. Also, tasks require different levels of annotation effort (eg. nuclei take longer to annotate than glands) and therefore exhaustively labelling data across tasks may not be feasible. As a result of the challenges above and to enable the effective utilisation of multiple independent datasets, in this paper we focus on MTL, rather than MLL.

\subsection{Multi-task learning for CPath}

Several recent CPath models utilise MTL to obtain a representative set of features for transfer learning. For example, Mormont \textit{et al.} \cite{mormont2020multi} considered 22 classification tasks within a simple MTL framework, where the learned weights are used to better initialise single task networks for transfer learning. Similarly, Tellez \textit{et al.} \cite{tellez2020extending} used MTL to optimise 4 tasks simultaneously and used the learned weights to encode a low-dimensional feature representation that generalised well to new tasks. Gamper \textit{et al.} \cite{gamper2021multiple} used MTL in combination with multiple instance captioning to learn representations from histopathology textbooks. Here, it was again shown that using MTL provided a strong feature encoder for transfer learning. Despite the demonstrated benefit of using MTL for transfer learning, subsequent training may still require a single model per task, which comes at an extra computational cost and demands additional training time.

 \begin{figure*}[t]
	\centering
    \includegraphics[width=0.99\textwidth]{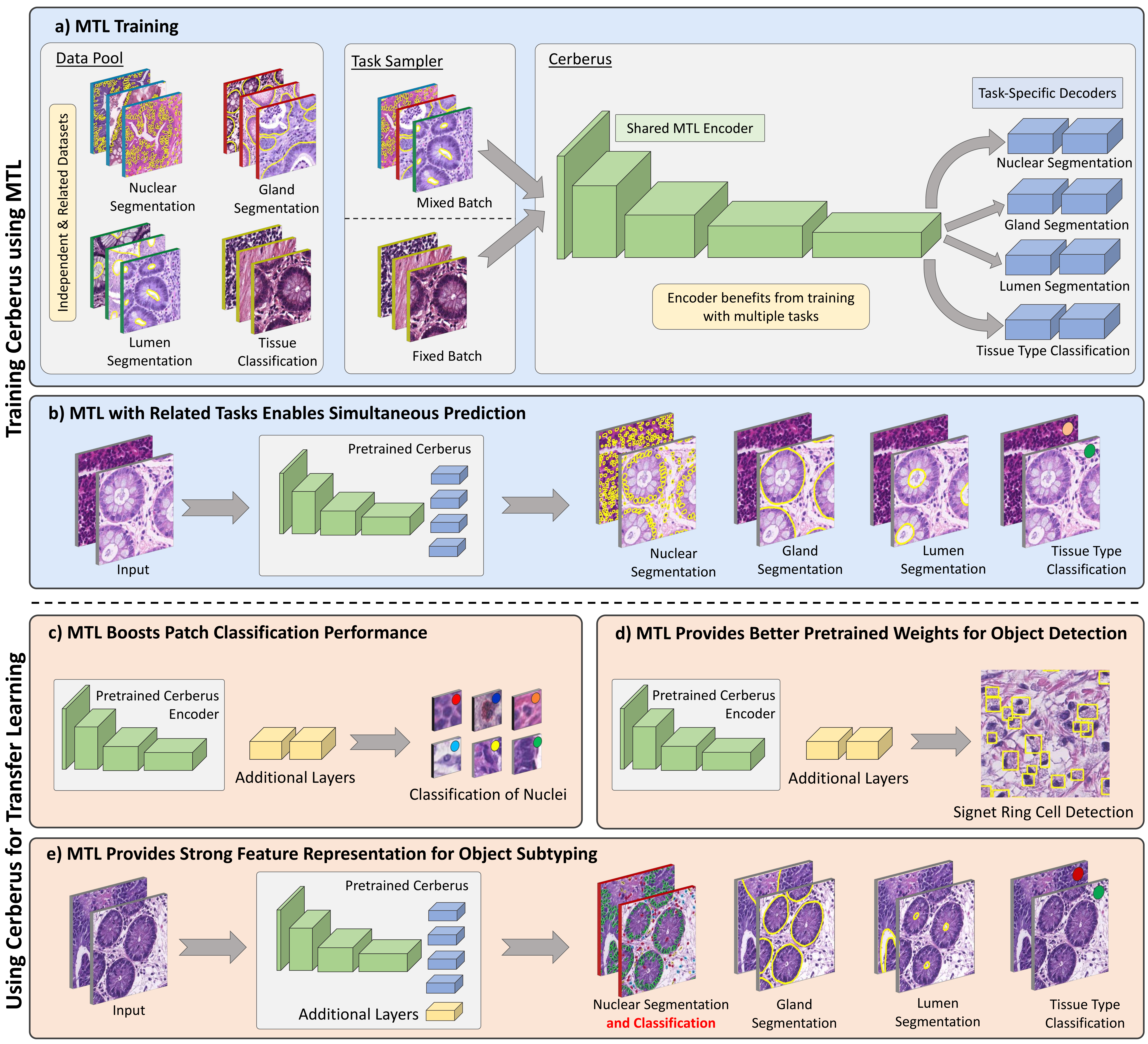}
	\caption{MTL training strategy (a), where patches are extracted from each task-level dataset by the task sampler and are fed into Cerberus. The sampler either chooses a fixed or mixed batch, where input samples are drawn from the same task or a mixture of tasks, respectively. Training with Cerberus (b) enables meaningful predictions to be made simultaneously. Using the trained Cerberus encoder can also help improve the performance of other tasks, such as patch classification (c), object detection (d) and object subtyping (e). For (c) and (d), only the pretrained Cerberus encoder is leveraged and additional classification or detection layers are added. For (e), the entire Cerberus architecture is used and extended by adding additional task-dependent decoders.}
	\label{fig:network}
\end{figure*}

Instead, it may also be desirable to perform well for all considered MTL tasks at the \textit{output} of the network. In this way, MTL provides a mechanism to learn from multiple independent datasets and enables simultaneous predictions to be made. However, this is only recommended if all considered tasks are aligned. For example, in work by Mormont \textit{et al.} \cite{mormont2020multi} one task uses images stained via immunohistochemistry (IHC) for identification of tumour and stroma regions in colorectal tissue, while another task uses Haematoxylin \& Eosin (H\&E) stained bone marrow tissue for cell classification. Therefore, unexpected results may be obtained at the output of the cell classification branch if presented with IHC stained data. This is because different input data assumptions exist for each task. In fact, none of the above-mentioned studies use MTL for the primary purpose of simultaneous prediction. With the use of different input assumptions, there may be task entanglement and consequently gradient conflicts \cite{liu2021conflict, sener2018multi} during network optimisation. This is a major reason why recent MTL methods in CPath can not match single-task performance at the network output \cite{gamper2020multi}.

\subsection{Supervised learning datasets for CPath}
MTL can help improve data efficiency because tasks with limited examples can benefit from the data provided by other tasks. Despite this, it is desirable to ensure that a sufficient number of examples from a variety of different sources are available per task, to increase the likelihood of strong generalisation. 

Over recent years, there has been a growing effort in generating labelled datasets for CPath \cite{graham2019hover,kumar2019multi, verma2020multi, amgad2021nucls, naylor2018segmentation, gamper2020pannuke, graham2021lizard}. Due to differences in the tissue type, resolution and stain, there may exist significant variation between these datasets. For example, Amgad \textit{et al.} \cite{amgad2019structured} generated a dataset of annotated regions in H\&E stained breast tissue at 20$\times$ magnification, whereas \cite{francesco_ciompi_2019_3513571} created a dataset of lymphocyte counts in tissue with IHC staining at 40$\times$ magnification. This variation makes it difficult to aggregate datasets that enable simultaneous prediction, where it is necessary for task-level datasets to have the same input assumptions. As well as this, manual annotation makes it difficult to generate large-scale datasets for segmentation tasks in computational pathology. Iterative labelling strategies \cite{tavolara2020segmentation,graham2021lizard,jaber2021deep}, which use a human-in-the-loop mechanism, have recently been used in the literature and can help significantly decrease the time required for annotation, while still ensuring a high level of accuracy.

If we assume that all data irrespective of the task should be H\&E stained, from colorectal tissue and at 20$\times$ objective magnification, then there exists a large amount of available data. Graham \textit{et al.} \cite{graham2021lizard} generated the largest dataset for nuclear instance segmentation and classification in CPath, consisting of 495,179 labelled nuclei from multiple sources. As part of the GlaS challenge, Sirinukunwattana \textit{et al.} \cite{sirinukunwattana2017gland} curated a dataset of 1,530 annotated gland boundaries originating from University Hospitals Coventry and Warwickshire (UHCW). This gland segmentation data was extended by Graham \textit{et al.} \cite{graham2019mild} into the CRAG dataset, with an additional 3,054 annotated glands from UHCW. As well as segmentation data in colorectal tissue, there exists available patch-based classification data that enables the development of further supervised learning methods. For example, Kather \textit{et al.} \cite{kather2019predicting} and Javed \textit{et al.} \cite{javed2020cellular} both created large datasets for tissue type classification consisting of 100 thousand and 280 thousand patches, respectively. As a result of the number of existing available datasets, we use H\&E stained images from colorectal tissue in all experiments within this paper. 

\section{Methods}
In this section, we describe our proposed MTL-based Cerberus model that enables the simultaneous prediction of aligned tasks in CPath. We report how the learned feature representation can be used to help improve the performance of additional tasks with transfer learning. We then describe the data generation strategy to enable the collection of large datasets for segmentation tasks in CPath. This strategy is used to significantly increase the amount of data used in experiments performed in Section \ref{section:results}.

\subsection{The \textit{Cerberus} model architecture}
Cerberus is a fully convolutional neural network with a shared encoder $\Phi$ and $T$ independent decoders $\Psi_t$, which make a prediction for each task $t$. Utilising a shared encoder ensures that a general representation is learned, where each task can benefit from features learned from other tasks. This can be especially useful when certain tasks do not have access to a large amount of data.

We use a ResNet34 encoder \cite{he2016deep} for ease of use of weights after training with MTL in downstream applications. For segmentation tasks, we use a U-Net \cite{ronneberger2015u} style decoder that incrementally upsamples the features by a factor of 2. After each upsampling operation, we incorporate features from the encoder with skip connections, followed by 2 convolutions (3$\times$3 kernel) with batch normalisation \cite{ioffe2015batch}. This is repeated until the features have the same spatial dimensions as the input. For patch classification tasks, we use global average pooling to reduce the features at the output of the encoder to a $k$ dimensional vector, which is then followed by 2 fully connected (FC) layers. In particular, we set $k$ to 256 and use Dropout \cite{srivastava2014dropout} between each FC layer, with a dropout rate of 0.3. An overview of the architecture can be seen in Figure \ref{fig:network}a. 

\subsection{Cerberus training}
Our overall MTL training strategy for Cerberus incorporates data from multiple independent sources and makes a simultaneous prediction at the output of the network. This training pipeline consists of the following steps: (\textit{i}) sampling patches from each task, (\textit{ii}) feeding patches through a convolutional neural network with a shared encoder and (\textit{iii}) task-level loss aggregation.

\subsubsection{Task sampler} 
\label{section:task-sampler}

Our overall dataset $D$ consists of $T$ task-dependent datasets $D_t$, which must be sampled appropriately during batch aggregation. This is because the network parameters are updated according to the data that is present in a particular batch. Careful consideration may be needed when there exist significant differences in the number of samples within each dataset $D_t$ or if certain tasks are more challenging than others. More information on optimisation will be provided in Section \ref{section:dynamic_training}.

When aggregating a batch, we need to ensure that all patches are the same size, so that they can be processed by the network. However, images for different tasks may have different spatial dimensions. It is possible to perform resizing or cropping to ensure that patches across all tasks have the same dimensions, but this may have an unwanted effect on the image quality or spatial context. Therefore, we group tasks into \textit{super tasks}, which can contain multiple sub-tasks with the same input dimensions. If considering multiple \textit{super tasks}, then one must be first randomly selected before batch aggregation. In our case, segmentation tasks (input dimensions 448$\times$448) and tissue type classification (input dimensions 144$\times$144) are treated as different \textit{super tasks}, which are selected with probabilities of 0.7 and 0.3 respectively.

In this work, we investigate the effect of using two different batch aggregation methods: \textit{mixed batch} and \textit{fixed batch} sampling. A mixed batch contains patches from multiple tasks, whereas a fixed batch contains data from a single task. For each patch within a mixed batch, we first select a task-level dataset $D_t$ from the entire multi-task dataset $D$ with a probability of $p_t$. Once a task is selected, a patch is randomly chosen from the corresponding dataset. This two-step process is repeated $n$ times, where $n$ is the number of samples in the batch. Rather than selecting a random task for each sample in a batch, for a \textit{fixed batch} we sample a task dataset $D_t$ once per batch and then select $n$ random samples from that given dataset. In all experiments, we set $p_t = \frac{1}{T}$.

In Figure \ref{fig:network}a, we show a simple example of the output of our task sampler, with a batch size of 3. Here, each coloured border denotes a separate task. 

\subsubsection{Dynamic training and loss aggregation}
\label{section:dynamic_training}
Previous work has shown that optimisation of MTL models can be difficult, especially when tasks are in conflict with each other \cite{lin2019pareto}. In CPath, this may involve using images from different tissue types or with different pixel resolution. This challenge can be exacerbated when each task has a different associated loss function, which makes balancing each task-level loss and consequently overall network optimisation quite challenging. To help ease the training of our model, we ensure that all considered tasks have the same input data assumptions \textbf{and} use a cross-entropy-based loss function to optimise the model. Cross-entropy loss is widely used for both segmentation and classification tasks and therefore is an appropriate choice if we want to use a similar loss function across different tasks. Note, when considering an instance segmentation task, we use weighted cross-entropy, as used in the original U-Net model \cite{ronneberger2015u}.

When training the network, we use dynamic weight freezing that is dependent on how batches are aggregated by the task sampler. After passing a batch through the network, the weights in the encoder $\Phi$ are always updated irrespective of how the batch is sampled, whereas the weights of decoder $\Psi_t$ are only updated when at least one example from task $t$ is present.

Hereby we define the loss function when we jointly train Cerberus:
\begin{equation}
\mathcal{L} = \sum_{t \in [1,T]}{\sum_{\rho \in D_t}\mathcal{L}_t(\{\Phi,\Psi_t\},x_{\rho},y_{\rho})}
\end{equation}
In this equation, $\rho \in D_t$ denotes a sample patch $\rho$ which belongs to the dataset $D_t$ of the task $t$. Meanwhile, $x_{\rho}$ and $y_{\rho}$ are predictions generated by the decoder $\Psi_t$ and the ground truth for the sample $\rho$ respectively. In practice, the loss of each task $t$ is computed by masking samples from other tasks with zeros. In other words, we multiply the resulting loss of all samples with a mask to set their subsequent gradient calculation to zero.

\subsection{Cerberus for transfer learning}

As a result of training Cerberus on a large amount of data, we can leverage the encoder to improve the performance of additional tasks with transfer learning. For example, the Cerberus feature representation can be used to enhance the performance of classification tasks or the model parameters can be used as pretrained weights for better model initialisation. The learned representation may also be used to extend the trained model and enable the subtyping of initially localised objects. To highlight the advantage of utilising Cerberus for transfer learning, we experiment with 3 different applications. First, we see whether the feature representation learned by Cerberus can improve both patch classification and object subtyping and then we see whether Cerberus can provide better pretrained weights for object detection.

An overview of each of the tasks mentioned above is provided in Figure \ref{fig:network}c-e. These are treated as single-task problems and therefore a simple batch aggregation strategy is used, rather than utilising our MTL task sampler, as described in Section \ref{section:task-sampler}.

\subsubsection{Patch classification}
To test the strength of the features learned by Cerberus, we use them for the task of subtyping small patches centred at individual objects, such as nuclei. In particular, we utilise the trained Cerberus encoder $\Phi$ and add several output layers to predict the corresponding category. For this purpose, we aggregate features from several positions in the encoder before applying global average pooling and two fully connected layers to obtain the prediction. We account for features at different scales within the encoder in order to demonstrate the feature strength at multiple points within the network. We also freeze the encoder weights and train the output layers in isolation to capture the contribution of the feature representation.

\subsubsection{Object subtyping}
\label{section:methods_subtyping}
To give further merit to the Cerberus feature representation, we show how it can be leveraged to sub-categorise each of the initially considered binary segmentation tasks. For this, we add an additional decoder $\Psi_{T+1}$ to Cerberus, dedicated to the task of subtyping a particular output, such as the nuclei. Then, we train $\Psi_{T+1}$ in isolation, while freezing both $\Phi$ and the decoders $\Psi_t$ $\forall$ $t \in [1, T]$. We treat this task as a pixel-based classification problem and therefore utilise the same decoder architecture used by the other segmentation tasks. After training the additional subtyping branch, Cerberus can predict all tasks simultaneously. 

Training the subtyping branch in isolation means that we do not have to cater for balancing the different components of the loss function between each task. Therefore, we use a combination of cross-entropy and Dice loss, which can also help counter class imbalance in the data \cite{graham2019hover}. Also, we only calculate the loss within foreground pixels by performing a masking operation on the loss function. We do this because the foreground pixels have already been localised by the associated binary segmentation branch and therefore differentiating between foreground and background is not needed. Concretely, the loss at the output of the subtyping branch $\Psi_{T+1}$ for input $X$ is defined as:

\begin{equation}
    \mathcal{L}_{T+1}(X) = \mathcal{L}_\alpha(X) + \mathcal{L}_\beta(X),
\end{equation}

\noindent where $\mathcal{L}_\alpha$ and $\mathcal{L}_\beta$ refer to the masked cross-entropy and Dice losses respectively. Specifically, $\mathcal{L}_\alpha(X)$ and $\mathcal{L}_\beta(X)$ are defined as follows:

\begin{equation}
    \mathcal{L}_\alpha(X) = - \sum_{k \in K}\frac{1}{|\nu_k|}\sum_{i \in \nu_k} y_{i,k}(X) \cdot \text{log}\hat{y}_{i,k}(X)
\end{equation}

\begin{equation}
    \mathcal{L}_\beta(X) = \sum_{k \in K} 1 - \frac{2 \cdot \sum_{i \in \nu_k}(y_{i,k}(X) + \hat{y}_{i,k}(X)) + \epsilon}{\sum_{i \in \nu_k}y_{i,k}(X) + \sum_{i \in \nu_k}\hat{y}_{i,k}(X) + \epsilon}.
\end{equation}

\noindent Here, $K$ is the number of classes, $y$ denotes the prediction and $\hat{y}$ is the ground truth. Therefore, $y_{i,k}(X)$ corresponds to the prediction for input image $X$ for class $k$ and pixel $i$. $\nu_k$ denotes the set of foreground pixels belonging to class $k$ and $\epsilon$ is a smoothing parameter used in the Dice loss to avoid division by zero. 

\subsubsection{Object detection}
Instead of adding additional decoders so that Cerberus performs an increasing number of tasks, we may want to utilise the weights in the trained Cerberus encoder for improved weight initialisation in a separate network. It is a common strategy to initialise networks with weights obtained from sampling from the normal distribution \cite{glorot2010understanding, he2015delving} or from initially training on large datasets, such as ImageNet. However, we demonstrate that utilising weights obtained from training Cerberus can provide better initialisation for certain tasks in CPath, leading to improved performance. To showcase this, we train RetinaNet \cite{lin2017focal}, with our pretrained Cerberus encoder as the backbone, for the task of object detection. RetinaNet is a region proposal network (RPN) \cite{ren2015faster} that performs object detection by predicting the bounding boxes within an input image. RPNs are designed to efficiently predict region proposals with a wide range of scales and aspect ratios, which are then further processed to predict the location of object bounding boxes. For this experiment, our goal is to understand whether Cerberus can provide better initial weights, rather than isolating the contribution of the feature representation. Therefore, \textit{all} weights in the network are optimised together.

\subsection{Collecting large datasets for segmentation}
\label{section:dataset_gen}
Manually annotating datasets for deep segmentation models is a very time-consuming procedure. Many deep learning models can provide an accurate segmentation, especially in trivial areas with a distinct morphology. Therefore, it may be a waste of effort to still rely on manual annotation in areas where deep learning models can perform well. Therefore, to enable the collection of a large amount of accurately labelled data we employ a pathologist-in-the-loop annotation collection pipeline, which is a simplified version of the strategy implemented by Graham \textit{et al.} \cite{graham2021lizard}. Our method consists of three main steps: (\textit{i}) initial segmentation model training with manually annotated data, (\textit{ii}) pathologist-in-the-loop refinement and retraining, (\textit{iii}) final manual refinement.

\begin{figure}[t]
	\centering
    \includegraphics[width=0.95\columnwidth]{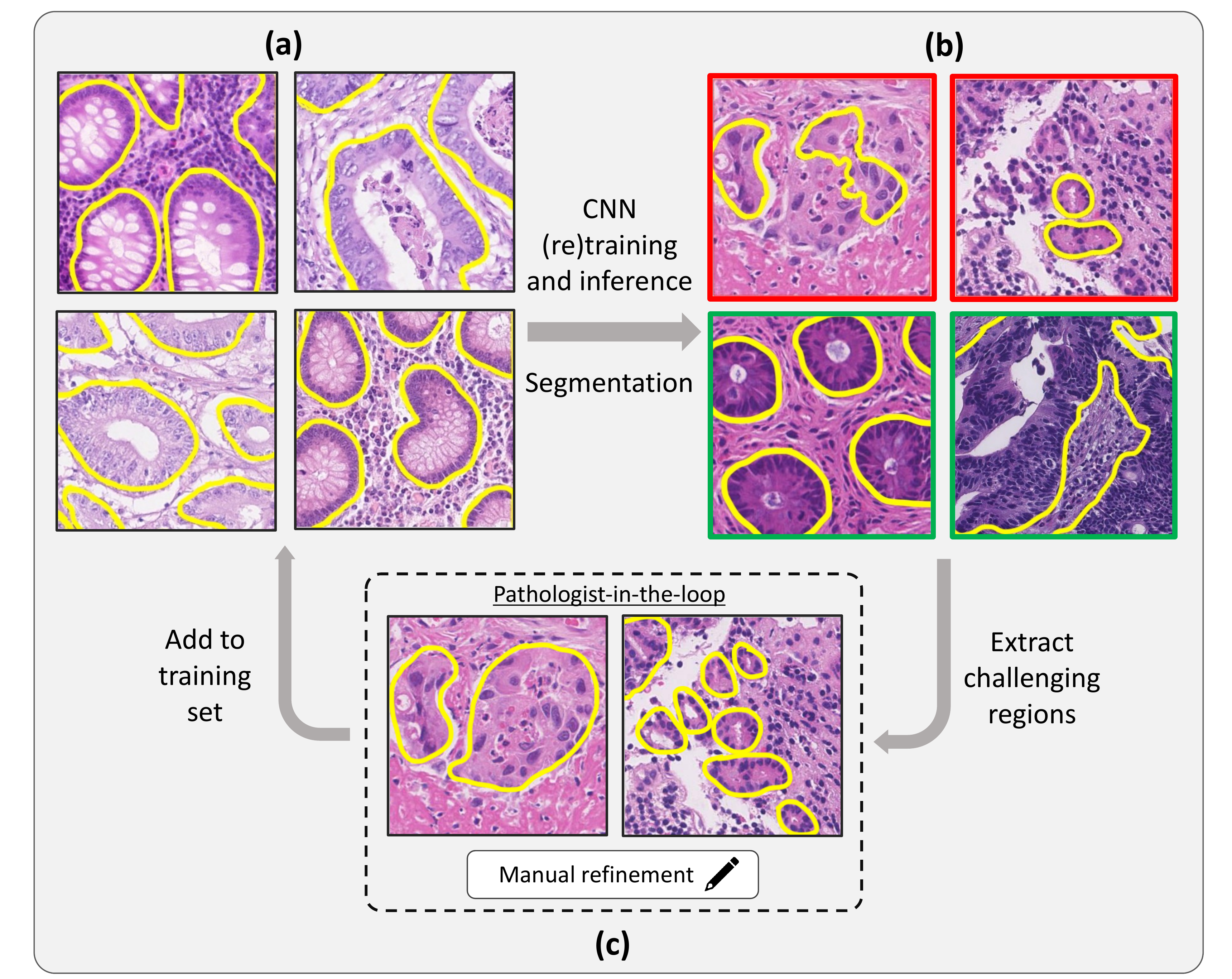}
	\caption{Annotating datasets at scale for segmentation: a) first train an initial segmentation model with available data, b) process unlabelled data and extract difficult regions where the model does not perform well and c) difficult regions are then refined and added to the training set before retraining the model. This sequence of steps is repeated until the data is accurately annotated. Difficult regions are shown with red borders in (b), otherwise they are shown with green borders.} 
	\label{fig:data_gen}
\end{figure}

First, we train an initial deep segmentation model on manually annotated data. This may be on existing available data if we are annotating objects with known publicly available datasets such as glands \cite{sirinukunwattana2017gland, graham2019mild}. Otherwise, if annotating an object with no available dataset in a given tissue type, then a manually annotated dataset needs to be created. 

After training the initial segmentation model, we process all data that we wish to annotate with the algorithm. With the input of a pathologist, we then extract image regions of size 1,000$\times$1,000 where the algorithm does not perform well and refine the results. These images, along with the corresponding refined annotations, are then added to the initial training set before re-training the model. This \textit{extract-refine-retrain} step is repeated until there is no noticeable improvement with the addition of further data. As a final step, we perform manual refinement of all data, which is then verified by a qualified pathologist. Segmented glands and lumina were presented for pathologist review when the data scientists were not sure about the visual quality of segmentation or noticed regions (glands or lumina) that could be potential false negatives. As a result, the number of actual glands/lumina reviewed and/or annotated by the pathologist was much less than the total number of glands/lumina. We provide an overview of our iterative labelling strategy in Figure \ref{fig:data_gen} and a summary of the number of annotated objects in Table \ref{table:dataset-summary}.

\section{Experiments and results}
\subsection{The datasets}

 \begin{figure*}[t]
\centering
\includegraphics[width=0.93\textwidth]{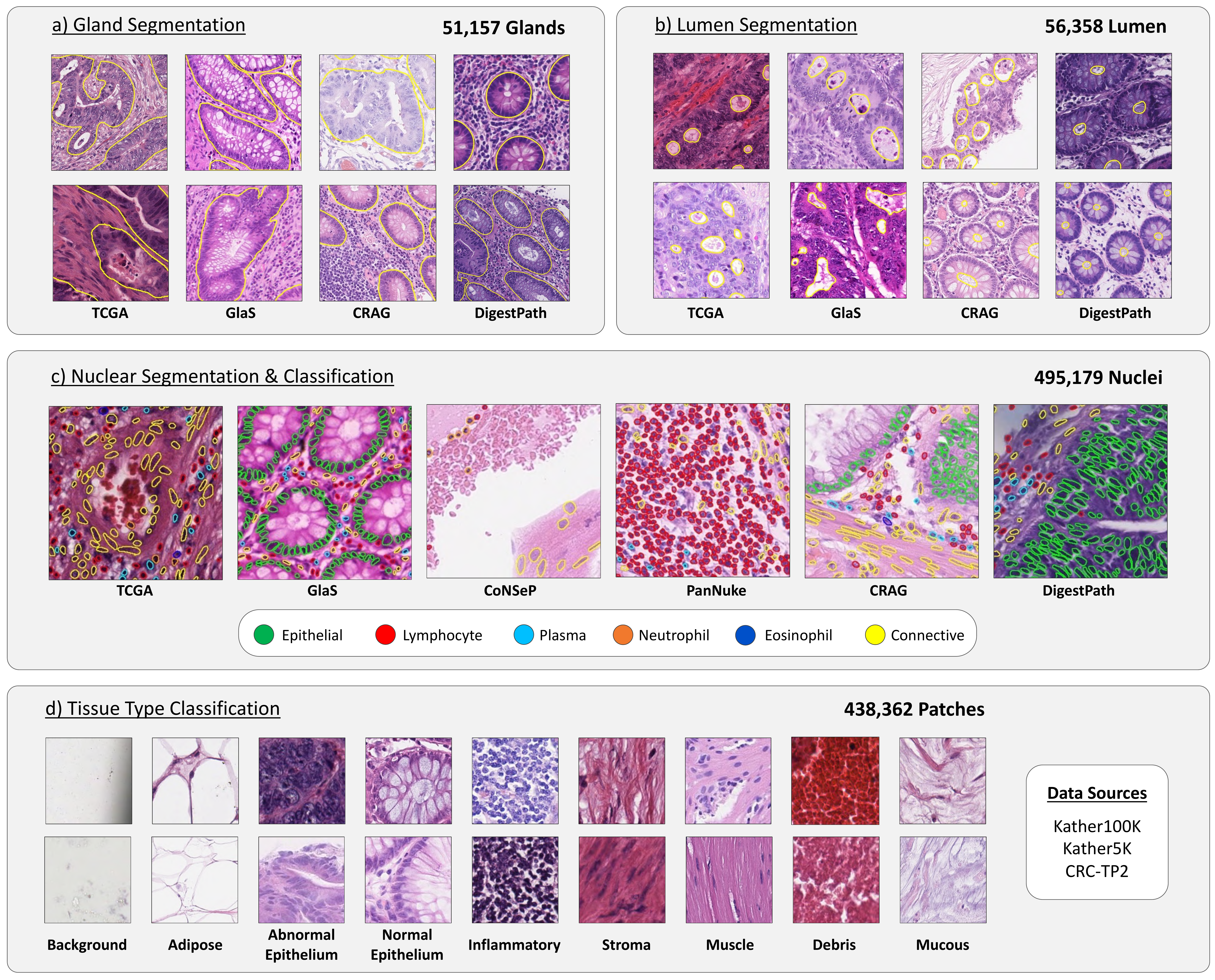}
\caption{Overview of the datasets used for a) gland segmentation, b) lumen segmentation, c) nuclear segmentation \& classification and d) tissue type classification. We also show the origin of the image regions used for each task.}
\label{fig:datasets}
\end{figure*}

In our MTL experiments, we utilise four task-level datasets. Specifically, we utilise a gland, lumen, nuclei and tissue type classification dataset, where each one contains image regions taken from H\&E stained colon tissue samples at 20$\times$ objective magnification (around 0.5$\mu$m/pixel).

\textbf{Gland dataset:} GlaS \cite{sirinukunwattana2017gland} and CRAG \cite{graham2019mild} are two of the most widely used available datasets for gland instance segmentation in colon histology images, containing a total of 1,602 and 3,209 annotated glandular boundaries, respectively. Despite GlaS and CRAG proving instrumental in the development of the recent automated gland segmentation models \cite{chen2017dcan, graham2019mild}, both datasets are from a single centre (UHCW) and therefore models trained on these may not generalise well to external data. Therefore, in addition, we annotate glands from two further datasets using our iterative pathologist-in-the-loop annotation method described in Section \ref{section:dataset_gen}. Specifically, we label 46,346 additional glands within images extracted from the DigestPath \cite{da2022digestpath} and TCGA datasets, respectively. 

\textbf{Lumen dataset:} There has been recent work on the collection of large annotated datasets for gland segmentation within colon tissue, but there are not many works on lumen segmentation, which can be beneficial for diagnosing certain conditions such as cribriform architecture and serrated polyps. To combat this, we annotate lumina within images taken from the GlaS, CRAG, DigestPath and TCGA datasets using our iterative labelling approach and accordingly use the data in associated experiments. For this, we manually label 6,863 lumen regions in GlaS and CRAG images as our initial annotated dataset and then use our iterative scheme to generate 49,495 lumen boundaries in the DigestPath and TCGA datasets. 

\textbf{Nuclei dataset:} We use the Lizard dataset \cite{graham2021lizard, graham2021conic}, which is the largest known instance segmentation and classification dataset in CPath containing nearly half a million labelled nuclei. Each nucleus is labelled according to the cell that it belongs to. Specifically, the labels used in the dataset are epithelial, lymphocyte, plasma, neutrophil, eosinophil and connective tissue. Here, the connective tissue is a broader category that includes endothelial cells, fibroblasts and muscle cells. When creating the dataset, image regions are extracted from the following 6 sources: CRAG \cite{graham2019mild}, GlaS \cite{sirinukunwattana2017gland}, CoNSeP \cite{graham2019hover}, PanNuke \cite{gamper2020pannuke}, DigestPath \cite{da2022digestpath} and TCGA. Therefore, Lizard is a diverse dataset and models trained on it may likely generalise to unseen examples. 

\textbf{Tissue type dataset:} We utilise 3 datasets for tissue type classification in colorectal images, that we refer to as CRC-TP2, Kather5K \cite{kather2016multi} and Kather100K \cite{kather2019predicting}, which contain 100,000, 5,000 and 333,362 patches respectively. Here, CRC-TP2 contains image patches from UHCW, whereas Kather5K and Kather100K contain patches from the pathology archive at the
University Medical Center Mannheim. CRC-TP2 is an extension of the dataset by Javed \textit{et al.} \cite{javed2020cellular}, where we add an extra 53,362 patches from an internal colon biopsy dataset \cite{wahab2021semantic}. Overall, the tissue types present in the dataset are background, adipose (fat), abnormal epithelium, normal epithelium, inflammatory, stroma, muscle, debris and mucous. Example image patches for each class can be seen in Figure \ref{fig:datasets}.

\textbf{Data split:} For segmentation tasks, we use TCGA data as an external test set and split the remaining data into 3 folds for cross-validation. For tissue type classification, we do not use an external test set but similarly split the data into 3 folds. When creating each split, we separate the data on a patient level to ensure that test data remains completely unseen. Then, for a given fold we assign each split as either a training, validation or testing subset. We also make sure that no patient overlap occurs between the subsets. For example, if a patch from fold 1 of the nuclei dataset belongs to a given patient, then that patient should not exist in any of the other datasets in folds 2 or 3. 

\begin{table}[t!]
\centering
\begin{tabular}{|c|c|c|c|}
\hline
& \textbf{Gland} & \textbf{Lumen} & \textbf{Nuclei}      \\ \hline
\textbf{Existing Data} & 4,811 & - & 495,179  \\ 
\textbf{Generated Data} & 46,346 & 56,358 & -  \\ 
\textbf{Total} & 51,157 & 56,358 & 495,179  \\ \hline 

\end{tabular}
\caption{Overview of the number of annotated objects used for segmentation tasks in this paper.}
\label{table:dataset-summary}
\end{table}

\subsection{Evaluation metrics}
\label{section:metrics}
\textbf{Segmentation:} We utilise the Dice score \cite{dice1945measures} and Panoptic Quality ($PQ$) \cite{kirillov2019panoptic} to measure the binary segmentation performance. Here, the Dice score measures how well a model can separate foreground pixels from the background and is defined as:

\begin{equation}
\text{Dice} = \frac{2 \times (|Y|\cap|\hat{Y}|)}{ |Y|+|\hat{Y}|},
\end{equation}

\noindent where $Y$ and $\hat{Y}$ are the prediction and ground truth maps respectively. However, the Dice score does not indicate whether neighbouring objects are correctly separated. Therefore, we also use $PQ$ to provide a measure for the quality of instance segmentation. $PQ$ matches predicted objects with ground truth objects based on whether their intersection-over-union ($IoU$) exceeds 0.5. This matching criterion is used to separate objects into true positives ($TP$), false positives ($FP$) and false negatives ($FN$). Then, $PQ$ is defined as:

\begin{equation}
\small
PQ = 
\underbrace{\frac{|TP|}{|TP|+\frac{1}{2}|FP|+\frac{1}{2}|FN|}}_{\text{Detection Quality(DQ)}}
\times
\underbrace{{\frac{\sum_{(y,\hat{y})\in{TP}}{IoU(y,\hat{y})}}{|TP|}}}_{\text{Segmentation Quality(SQ)}},
\end{equation}

 \noindent where $\hat{y}$ denotes a ground truth instance and $y$ denotes a predicted instance. We calculate $PQ$ individually for each image and then average the results to get the overall statistic. 
 
 When assessing multi-class instance segmentation, we calculate $PQ$ independently for each class and then report the average score. In previous work \cite{gamper2020pannuke, graham2021lizard}, $mPQ$ was defined by calculating $PQ$ for each image \textit{and} for each class before the overall results were averaged. However, the result of a particular image was not considered in the class average if that class was not present in the ground truth. To help avoid skipping images, we also report $mPQ^+$ \cite{graham2021conic}, which aggregates the statistics over \textbf{all} images before taking the class average.

\begin{table*}[t!]
\small
\centering
\begin{tabular}{|l|c|c|c|c|c|c|}
\hline
& \multicolumn{3}{c|}{\textbf{Cross Validation}} & \multicolumn{3}{c|}{\textbf{External Test}} \\
\cline{2-7} \textbf{Model} & \textbf{Binary PQ} & \textbf{mPQ} &  \textbf{mPQ$^+$}  & \textbf{Binary PQ} & \textbf{mPQ} &  \textbf{mPQ$^+$}\\ \hline
U-Net  & 0.564 $\pm$ 0.020 & 0.275 $\pm$ 0.014 & 0.324 $\pm$ 0.018 & 0.493 $\pm$ 0.051 & 0.235 $\pm$ 0.018 & 0.258 $\pm$ 0.034   \\ 
HoVer-Net & 0.583 $\pm$ 0.014 & 0.295 $\pm$ 0.018 & 0.409 $\pm$ 0.027 & 0.514 $\pm$ 0.026 & 0.285 $\pm$ 0.018 & 0.335 $\pm$ 0.017 \\ \hdashline
Cerberus & 0.588 $\pm$ 0.008 & 0.323 $\pm$ 0.020 & 0.396 $\pm$ 0.022 & 0.560 $\pm$ 0.028 & 0.308 $\pm$ 0.015 & 0.353 $\pm$ 0.009 \\ 
Cerberus$^+$ & \textbf{0.612 $\pm$ 0.010} & \textbf{0.358 $\pm$ 0.011} & \textbf{0.425 $\pm$ 0.019} & \textbf{0.568 $\pm$ 0.009} & \textbf{0.332 $\pm$ 0.011} & \textbf{0.388 $\pm$ 0.003} \\ 
\hline
\end{tabular}
\caption{Simultaneous nuclear segmentation and classification results of our proposed approach compared to recent state-of-the-art methods. Cerberus predicts the eroded instance target, whereas Cerberus$^+$ also predicts the object boundary.}
\label{table:compare-sota1}
\end{table*}

\begin{table*}[t!]
\small
\centering
\begin{tabular}{|l|c|c|c|c|c|c|}
\hline
& \multicolumn{2}{c|}{\textbf{Gland}} & \multicolumn{2}{c|}{\textbf{Lumen}} \\
\cline{2-5} \textbf{Model} & \textbf{Cross Val} & \textbf{External Test}  & \textbf{Cross Val} & \textbf{External Test} \\ \hline
U-Net  & 0.622 $\pm$ 0.030 & 0.459 $\pm$ 0.024 & 0.501 $\pm$ 0.008 & 0.240 $\pm$ 0.109 \\ 
MILD-Net & 0.647 $\pm$ 0.046 & 0.526 $\pm$ 0.027 & 0.522 $\pm$ 0.034 & 0.353 $\pm$ 0.088 \\
Rota-Net & 0.662 $\pm$ 0.048 & 0.573 $\pm$ 0.034 & 0.569 $\pm$ 0.042 & 0.436 $\pm$ 0.054 \\ \hdashline
Cerberus & \textbf{0.677 $\pm$ 0.028} & 0.640 $\pm$ 0.012 & 0.589 $\pm$ 0.006 & 0.525 $\pm$ 0.027 \\ 
Cerberus$^+$ & 0.674 $\pm$ 0.021 & \textbf{0.650 $\pm$ 0.004} & \textbf{0.590 $\pm$ 0.018} & \textbf{0.530 $\pm$ 0.005} \\ 
\hline
\end{tabular}
\caption{Gland and lumen segmentation results of our proposed approach compared to recent top-performing methods. Cerberus predicts the eroded instance target, whereas Cerberus$^+$ also predicts the object boundary.}
\label{table:compare-sota2}
\end{table*}

\textbf{Classification:} To assess the classification performance, we utilise mean average precision ($mAP$) and mean $F_1$ ($mF_1$). For both metrics, either average precision ($AP$) or $F_1$ is calculated and the average over the classes is reported to give $mAP$ and $mF_1$. $AP$ and $F_1$ are defined as:

\begin{equation}
\begin{aligned}
Pr_n &= \frac{TP_n}{TP_n + FP_n} \\
Re_n &= \frac{TP_n}{TP_n + FN_n} \\
\end{aligned}
\end{equation}

\begin{equation}
\begin{aligned}
AP &= \sum_n{(Re_n-Re_{n-1})Pr_n}
\end{aligned}
\end{equation}

\begin{equation}
\begin{aligned}
F^n_1 &= \frac{Pr_n * Re_n}{Pr_n + Re_n}
\end{aligned}
\end{equation}

\noindent
Upon using a threshold $n$ to convert a probability to a binary label, the predictions and ground truth are split into true positive ($TP$), false positive ($FP$) and false negative ($FN$) sets. The number of items in each of these sets is then utilised to calculate the precision ($Pr$) and recall ($Re$) values. Depending on the use case, $n$ may have a different meaning. For $AP$, $n$ denotes the $n$-th threshold level that is associated with a specific threshold value. For $n-1$, this extends to be the previous threshold level that is smaller than $n$. However, for $F_1$, we set the prediction label to be the one with the highest probability, rather than using a specific threshold value.

\subsection{Results}
\label{section:results}
 \subsubsection{Cerberus vs state-of-the-art approaches}
\label{section:cerberus-results}
First, we compare the segmentation performance of Cerberus with recent state-of-the-art approaches in CPath and provide the results in Tables \ref{table:compare-sota1} and \ref{table:compare-sota2}. Here, we only consider MTL with mixed batch sampling, due to its superior performance as shown in Table \ref{table:mtl-seg-results}. For this, we assess the binary segmentation performance of nuclei, glands and lumina using $PQ$ and then quantify the multi-class segmentation performance of nuclei using both $mPQ$ and $mPQ^+$. 

For nuclei, we compare the performance of our proposed approach with U-Net \cite{ronneberger2015u} and HoVer-Net \cite{graham2019hover}. HoVer-Net is a recent top-performing method for combined nuclear instance segmentation and classification \cite{verma2020multi, gamper2020pannuke} and therefore serves as a strong benchmark. To enable full comparison amongst all models for nuclear segmentation and classification, we add extra layers to the output of U-Net for nuclear subtyping. For both gland and lumen segmentation, we compare the performance of Cerberus with U-Net, MILD-Net \cite{graham2019mild} and Rota-Net \cite{graham2019rota}. MILD-Net and Rota-Net are competitive methods for both of these tasks and therefore serve as strong baselines. Here, Rota-Net segments the glands and lumina simultaneously, like in the original publication. All Cerberus and U-Net models use a simple 2-class eroded instance target, whereas MILD-Net and Rota-Net also incorporate the object boundary as an additional class. HoVer-Net uses a more complex instance segmentation target, consisting of horizontal and vertical distance maps. Therefore, if Cerberus can outperform competing methods using a simple instance segmentation target, then this holds great promise for achieving even better results when adding more complex targets. As an extra experiment, we see whether we can further boost the performance of Cerberus by also incorporating the object boundary. We denote this as Cerberus$^+$ in Tables \ref{table:compare-sota1} and \ref{table:compare-sota2}. For all U-Net models, we use batch normalisation, which was not used in the original publication \cite{ronneberger2015u}.

In Table \ref{table:compare-sota1}, we see that Cerberus outperforms both U-Net and HoVer-Net on nearly all measures for both binary and multi-class nuclear instance segmentation. This performance is further improved when using the object boundary as an additional target of the network and leads to significantly better performance for all measures compared to both U-Net and HoVer-Net. It may be worth noting that we report slightly different results than the HoVer-Net results reported by Graham \textit{et al.} \cite{graham2021lizard} because we ensure that the exact same input patches are used for a fair comparison. It is interesting to see that the boost in performance is more pronounced for the external test set, suggesting that MTL can help with generalisation. For gland and lumen segmentation, we observe from Table \ref{table:compare-sota2} that Cerberus exceeds the performance of U-Net, MILD-Net and Rota-Net on all measures. Again, this is especially prominent on the external test set. Despite the increase in performance not being large, we see a slight increase in performance when considering the object boundary, especially on the completely unseen data source. Therefore, this justifies predicting the object boundary for instance segmentation in all segmentation tasks considered by our network.

\begin{table*}[t!]
\small
\begin{subtable}{1\textwidth}
\sisetup{} 
\centering
\begin{tabular}{|c|c|c|c|c|c|c|c|c|}
\hline
& & & \multicolumn{2}{c|}{\textbf{Nuclei}} & \multicolumn{2}{c|}{\textbf{Gland}} & \multicolumn{2}{c|}{\textbf{Lumen}} \\ \cline{4-9}
 & \textbf{Task} & \textbf{Sampler} & \textbf{Dice} & \textbf{PQ} & \textbf{Dice} & \textbf{PQ} & \textbf{Dice} & \textbf{PQ}  \\ \hline
\parbox[t]{2mm}{\multirow{3}{*}{\rotatebox[origin=c]{90}{\textbf{U-Net}}}} & STL & Fixed & 0.767 $\pm$ 0.013 & 0.566 $\pm$ 0.021 & 0.891 $\pm$ 0.013 & 0.622 $\pm$ 0.030 & 0.659 $\pm$ 0.014 & 0.501 $\pm$ 0.008 \\ 
& MTL & Fixed  & 0.759 $\pm$ 0.004 & 0.560 $\pm$ 0.006 & 0.894 $\pm$ 0.006 & 0.636 $\pm$ 0.025 & 0.688 $\pm$ 0.017 & 0.536 $\pm$ 0.022  \\ 
& MTL & Mixed  & 0.762 $\pm$ 0.007 & 0.558 $\pm$ 0.011 & 0.897 $\pm$ 0.011 & 0.628 $\pm$ 0.015 & 0.697 $\pm$ 0.003 & 0.543 $\pm$ 0.011 \\ \hline
\parbox[t]{2mm}{\multirow{5}{*}{\rotatebox[origin=c]{90}{\textbf{ResNet-34}}}} & STL & Fixed  & 0.772 $\pm$ 0.002 & 0.581 $\pm$ 0.006 & 0.902 $\pm$ 0.007 & 0.669 $\pm$ 0.014 & 0.666 $\pm$ 0.060 & 0.508 $\pm$ 0.091  \\ \
& MTL & Fixed & 0.769 $\pm$ 0.007 & 0.572 $\pm$ 0.009 & 0.903 $\pm$ 0.001 & 0.677 $\pm$ 0.022 & 0.726 $\pm$ 0.015 & 0.585 $\pm$ 0.013 \\ 
& MTL & Mixed & 0.770 $\pm$ 0.003 & 0.574 $\pm$ 0.005 & 0.905 $\pm$ 0.005 & 0.674 $\pm$ 0.016 & 0.721 $\pm$ 0.018 & 0.590 $\pm$ 0.014  \\ \cdashline{2-9}
& IN-MTL & Mixed & \textbf{0.779 $\pm$ 0.004} & 0.588 $\pm$ 0.009 & 0.907 $\pm$ 0.002 & \textbf{0.683 $\pm$ 0.013} & 0.733 $\pm$ 0.019 & \textbf{0.592 $\pm$ 0.019}  \\
& IN-MTL+PC & Mixed & 0.778 $\pm$ 0.006 & \textbf{0.589 $\pm$ 0.008} & \textbf{0.909 $\pm$ 0.008} & 0.677 $\pm$ 0.028 & \textbf{0.735 $\pm$ 0.007} & 0.589 $\pm$ 0.006  \\ \hline
\end{tabular}
\caption{Cross validation results.}
\end{subtable}

\bigskip
\begin{subtable}{1\textwidth}
\sisetup{} 
\centering
\begin{tabular}{|c|c|c|c|c|c|c|c|c|}
\hline
& & & \multicolumn{2}{c|}{\textbf{Nuclei}} & \multicolumn{2}{c|}{\textbf{Gland}} & \multicolumn{2}{c|}{\textbf{Lumen}} \\ \cline{4-9}
 & \textbf{Task} & \textbf{Sampler} & \textbf{Dice} & \textbf{PQ} & \textbf{Dice} & \textbf{PQ} & \textbf{Dice} & \textbf{PQ}  \\ \hline
\parbox[t]{2mm}{\multirow{3}{*}{\rotatebox[origin=c]{90}{\textbf{U-Net}}}} & STL & Fixed & 0.713 $\pm$ 0.067 & 0.499 $\pm$ 0.058 & 0.741 $\pm$ 0.012 & 0.459 $\pm$ 0.024 & 0.281 $\pm$ 0.089 & 0.240 $\pm$ 0.109  \\  
& MTL & Fixed & 0.708 $\pm$ 0.061 & 0.486 $\pm$ 0.058 & 0.770 $\pm$ 0.100 & 0.471 $\pm$ 0.077 & 0.420 $\pm$ 0.051 & 0.376 $\pm$ 0.063  \\   
& MTL & Mixed & 0.738 $\pm$ 0.017 & 0.505 $\pm$ 0.017 & 0.824 $\pm$ 0.061 & 0.492 $\pm$ 0.039 & 0.446 $\pm$ 0.063 & 0.377 $\pm$ 0.059  \\  \hline
\parbox[t]{2mm}{\multirow{5}{*}{\rotatebox[origin=c]{90}{\textbf{ResNet-34}}}} & STL & Fixed  & 0.758 $\pm$ 0.021 & 0.554 $\pm$ 0.030 & 0.820 $\pm$ 0.104 & 0.566 $\pm$ 0.082 & 0.388 $\pm$ 0.016 & 0.313 $\pm$ 0.041  \\ 
& MTL & Fixed & 0.739 $\pm$ 0.032 & 0.527 $\pm$ 0.042 & 0.864 $\pm$ 0.031 & 0.602 $\pm$ 0.020 & 0.594 $\pm$ 0.046 & 0.470 $\pm$ 0.010  \\ 
& MTL & Mixed & 0.742 $\pm$ 0.027 & 0.531 $\pm$ 0.038 & 0.881 $\pm$ 0.013 & 0.617 $\pm$ 0.004 & 0.622 $\pm$ 0.053 & 0.492 $\pm$ 0.013  \\  \cdashline{2-9}
& IN-MTL & Mixed & 0.773 $\pm$ 0.031 & \textbf{0.560 $\pm$ 0.037} & 0.875 $\pm$ 0.072 & 0.619 $\pm$ 0.064 & 0.637 $\pm$ 0.057 & 0.516 $\pm$ 0.056  \\  
& IN-MTL+PC & Mixed & \textbf{0.774 $\pm$ 0.024} & \textbf{0.560 $\pm$ 0.028} & \textbf{0.908 $\pm$ 0.010} & \textbf{0.640 $\pm$ 0.012} & \textbf{0.666 $\pm$ 0.014} & \textbf{0.525 $\pm$ 0.027}  \\  \hline
\end{tabular}
\caption{External test results.}
\end{subtable}

\caption{Comparison of results obtained for binary segmentation using MTL and STL. Sampler indicates how patches are aggregated into batches. A \textit{Fixed} batch contains patches from a single task and a \textit{Mixed} batch contains patches from multiple tasks (Section \ref{section:task-sampler}). IN refers to ImageNet pretrained models and PC denotes patch classification. When using MTL with PC, a batch contains either only patches for segmentation or PC (Section \ref{section:task-sampler}).}
\label{table:mtl-seg-results}
\end{table*}

\begin{table}[t!]
\small
\centering
\begin{tabular}{|l|c|c|}
\hline
\textbf{Task} & \textbf{mAP} & \textbf{mF$_1$} \\ \hline
IN-STL  & 0.949 $\pm$ 0.002 & 0.885 $\pm$ 0.001 \\
IN-MTL+PC & 0.948 $\pm$ 0.006 & 0.883 $\pm$ 0.007 \\
\hline
\end{tabular}
\caption{Comparison of performance with models trained using multi-task and single-task learning for tissue type classification.}    
\label{table:patch-tissue-class}
\end{table}

\subsubsection{Multi-task learning vs single-task learning for segmentation}
\label{section:mtl-vs-stl-results}
We perform extensive experiments to fully understand the contribution of multi-task learning for segmentation. For this experiment, we only consider the binary output for segmentation tasks and consider the eroded map as the instance target for simplicity. We report results for further subtyping of the segmentation output in Section \ref{section:mtl-nuclei-results}. In particular, we compare single-task learning (STL) methods, where a devoted network is trained individually per task, with multi-task networks, where tasks are optimised simultaneously. We also experiment with two different sampling strategies for MTL: (\textit{i}) sampling a single task per batch and (\textit{ii}) sampling a mixture of tasks per batch, referred to as \textit{fixed} and \textit{mixed} batches, respectively in \ref{section:task-sampler}. We investigate the above strategies using encoder-decoder based model architectures with hard parameter sharing. Specifically, we experiment with both a basic U-Net style encoder \cite{ronneberger2015u} with batch normalisation and a ResNet34 encoder \cite{he2016deep}. We assess the performance of multiple models to remove any bias from the model architecture and instead solely focus on the contribution of the training scheme. After gaining insight into the effect of MTL for segmentation tasks, we investigate whether ImageNet pretrained weights and utilising patch classification as an auxiliary task can further increase the segmentation performance. As described in Section \ref{section:task-sampler}, a batch must contain a single \textit{super task}, which needs to be initially selected if considering both segmentation and patch classification.

Our major aim is that we want the performance of networks trained using MTL to be \textbf{at least as good} as those trained using STL. This would reduce the computational requirements during training and inference because just a single network could be used to perform tasks simultaneously. We can see from Table \ref{table:mtl-seg-results} that in fact we achieve superior performance for most tasks considered during our experiments. In particular, we observe that both U-Net and ResNet34 alternatives benefit from MTL, especially for gland and lumen segmentation. This observation is more pronounced on the external test set, suggesting that the features learned during MTL can help with generalisation. The greatest improvement from training with MTL is observed for lumen segmentation, especially on the external test set. This can potentially be attributed to learning complementary information regarding the location of glandular boundaries. Despite not observing an obvious increase in performance for nuclear segmentation, results are comparable, suggesting that we do not require a separate network to perform this task. We suspect that this may be because we are already using a very large dataset for nuclear segmentation and so the addition of complementary tasks may not be particularly advantageous.

Although we achieve similar performance during cross-validation, it can be seen that utilising a mixed batch leads to significantly better performance on the external test set. Therefore, for the remainder of the experiments in this paper, we only consider MTL with mixed batch sampling. We also hypothesise that mixed batch sampling will scale better when using a large number of tasks because for a given batch, there is a greater probability of at least one sample being present for a particular task. However, this assumption only holds if the batch size is greater than the number of tasks. Further investigation into the relationship between batch types (mixed/fixed) and batch size may be required in future work to fully understand what kind of batches work best for MTL.

For all tasks, we observe that using a ResNet encoder results in a significantly better performance compared to using a conventional U-Net approach. We also observe that initialising the ResNet encoder with ImageNet weights, rather than randomly, gives a further boost in the performance. From Table \ref{table:mtl-seg-results} we can see that with the addition of patch classification, results are comparable for cross-validation, but have improved on the external test set. This suggests that utilising the additional data from the patch classification task leads to the shared encoder learning more representative features, and hence leads to better performance on external data. Of course, as an additional benefit, the network can predict the tissue type at the output of the network, which can potentially lead to more powerful downstream pipelines. It must be noted that results using ImageNet-pretrained MTL models were obtained to assess potential performance gains over MTL. Therefore, they should not be compared directly with STL models. We report the patch classification performance in Table \ref{table:patch-tissue-class}, where we observe that the multi-task and single-task models achieve similar performance. Therefore, performance is not compromised when considering multiple tasks alongside patch classification. For this experiment, we compare our approach with STL pretrained with ImageNet to provide a fair comparison.

\begin{table*}
\small
\centering
\begin{tabular}{|c|c|c|c|c|c|c|}
\hline
 & & \multicolumn{2}{c|}{\textbf{Cross Validation}} & \multicolumn{2}{c|}{\textbf{External Test}} \\
 \cline{3-7}
 
 \textbf{Features} & \textbf{Mode}   & \textbf{mAP} & \textbf{mF$_1$} & \textbf{mAP} & \textbf{mF$_1$} \\ \hline
IN & Patch & 0.535 $\pm$ 0.018 & 0.491 $\pm$ 0.026 & 0.385 $\pm$ 0.013 & 0.355 $\pm$ 0.020 \\ 
SimCLR & Patch  & 0.508 $\pm$ 0.005 & 0.464 $\pm$ 0.014 & 0.432 $\pm$ 0.029 & 0.403 $\pm$ 0.026 \\
SimCLR$^+$ & Patch  & 0.592 $\pm$ 0.012 & 0.557 $\pm$ 0.012 & 0.488 $\pm$ 0.023 & 0.442 $\pm$ 0.009 \\
STL Nuclei & Patch & 0.635 $\pm$ 0.033 & 0.588 $\pm$ 0.029 & 0.572 $\pm$ 0.022 & 0.538 $\pm$ 0.020  \\  
STL Gland & Patch & 0.542 $\pm$ 0.027 & 0.499 $\pm$ 0.041 & 0.495 $\pm$ 0.042 & 0.456 $\pm$ 0.056  \\  
STL Lumen & Patch & 0.520 $\pm$ 0.022 & 0.478 $\pm$ 0.038 & 0.452 $\pm$ 0.059 & 0.418 $\pm$ 0.073  \\  
MTL  & Patch & 0.587 $\pm$ 0.032 & 0.543 $\pm$ 0.039 & 0.558 $\pm$ 0.005 & 0.506 $\pm$ 0.010  \\ 
IN-MTL & Patch &  0.607 $\pm$ 0.033 & 0.562 $\pm$ 0.038 & 0.540 $\pm$ 0.035 & 0.492 $\pm$ 0.041   \\
IN-MTL+PC & Patch &  \textbf{0.651 $\pm$ 0.067} & \textbf{0.601 $\pm$ 0.049} & 0.573 $\pm$ 0.013 & \textbf{0.540 $\pm$ 0.019} \\\hdashline
IN-MTL+PC$^*$ & Patch &  0.628 $\pm$ 0.030 & 0.577 $\pm$ 0.038 & \textbf{0.586 $\pm$ 0.024} & 0.518 $\pm$ 0.034  \\ 
Ciga \textit{et al.} \cite{ciga2022self} & Patch &  0.615 $\pm$ 0.004 & 0.566 $\pm$ 0.019 & 0.540 $\pm$ 0.035 & 0.492 $\pm$ 0.041  \\\hline 
\end{tabular}
\caption{Comparison of different deep features for nuclear classification. All models consider a patch of size 32$\times$32 extracted at the centre of each nucleus. We only consider MTL with \textit{mixed batch} sampling due to its superior performance in Table \ref{table:mtl-seg-results}.  For all models, we freeze the ResNet weights and only train the classification layers.  Models above the dashed line use a ResNet34 backbone,  whereas models below the dashed line use ResNet18 to enable cross-comparison with Ciga \textit{et al.} \cite{ciga2022self}, which uses SimCLR. The difference between this and SimCLR/SimCLR$^+$ reported on rows 2 and 3 is the data used for training. Here, SimCLR uses the same data used during MTL and SimCLR$^+$ uses over 2 million colon H\&E image patches from many datasets.}
\label{table:feat-class-results}
\end{table*}

\begin{table*}
\small
\centering
\begin{tabular}{|c|c|c|c|c|c|c|c|}
\hline
 &  & & \multicolumn{2}{c|}{\textbf{Cross Validation}}  &  \multicolumn{2}{c|}{\textbf{External Test}} \\ \cline{4-8}
 \textbf{Features} & \textbf{Mode} & \textbf{Loss} & \textbf{mAP} & \textbf{mF$_1$} & \textbf{mAP} & \textbf{mF$_1$} \\ \hline
IN-MTL+PC & Pixel$_A$ &  CE  & 0.493 $\pm$ 0.021 & 0.477 $\pm$ 0.013 & 0.447 $\pm$ 0.025 & 0.432 $\pm$ 0.012 \\
IN-MTL+PC & Pixel$_A$ & CE \& Dice & 0.537 $\pm$ 0.017 & 0.545 $\pm$ 0.015 & 0.494 $\pm$ 0.030 & 0.497 $\pm$ 0.012 \\
IN-MTL+PC & Pixel$_A$ &  Masked CE \& Dice & 0.576 $\pm$ 0.012 & 0.566 $\pm$ 0.008 & 0.529 $\pm$ 0.032 & 0.509 $\pm$ 0.020 \\
IN-MTL+PC & Pixel$_B$ & Masked CE \& Dice & \textbf{0.678 $\pm$ 0.035} & \textbf{0.710 $\pm$ 0.030} & \textbf{0.634 $\pm$ 0.013} & \textbf{0.669 $\pm$ 0.003} \\
\hline 
\end{tabular}
\caption{Results for nuclear classification with different loss strategies by making a prediction per pixel.  Pixel$_A$ performs nuclear classification by adding an extra classification layer at the output of the nuclear segmentation branch, whereas Pixel$_B$ uses a devoted upsampling branch.}
\label{table:feat-class-seg-results}
\end{table*}

\subsubsection{Cerberus feature representation for patch-based nuclear classification}
\label{section:mtl-nuclei-results}
After training Cerberus, our next experiment involves understanding whether the learned features can be used to improve the performance of patch-level nuclear classification. We first assess whether the representation captured from training Cerberus is superior to that obtained from training on ImageNet in a supervised setting. Using ImageNet-trained networks is a common strategy to obtain a deep representation in the field of computer vision and therefore serves as a good benchmark for this experiment. Then, we compare single-task and self-supervised learning alternatives with histology-specific data. In particular, contrastive learning is a self-supervised approach that has shown great promise, without requiring labelled data. Therefore, we compare Cerberus features with those obtained using a recent contrastive learning method called SimCLR \cite{chen2020simple} on our combined gland, lumen and nuclei datasets. However, this may not fully unlock the potential of self-supervised methods that are only bounded by the amount of available image data. Therefore, we also add results using SimCLR trained with over 2 million H\&E colon input patches at 20$\times$ objective magnification and denote it as SimCLR$^+$. For this, in addition to the data that was used for MTL, we sampled many image patches from the PAIP challenge\footnote{The PAIP dataset is available to download here: \url{https://paip2020.grand-challenge.org/}} and TCGA datasets. As an additional benchmark, we also utilise a model provided by Ciga \textit{et al.} \cite{ciga2022self} that was optimised by training SimCLR on 57 histology datasets.

In Table \ref{table:feat-class-results}, we show results for nuclear classification using different feature representations. Each representation is obtained by passing images through a pretrained ResNet encoder, where weights were initially obtained using one of the strategies described above. Features are passed to several trainable output layers to predict the nuclear category. We aggregate features from several positions of the encoder before applying global average pooling and two fully connected layers to obtain the final prediction. For most entries in Table \ref{table:feat-class-results} above the dashed line, we utilise a ResNet34 model for feature extraction. However, Ciga \textit{et al.} \cite{ciga2022self} do not have this model available and instead provides a ResNet18 alternative. Therefore, to enable a like-for-like comparison, we also retrain Cerberus with a ResNet18 backbone and compare the corresponding results.

We extract a dataset of patches of size 32$\times$32 pixels centred at each nucleus within our nuclear segmentation dataset and use the same splits that were initially used for feature extraction to ensure that all test data remains completely unseen. We observe from Table \ref{table:feat-class-results} that all features obtained using single or multi-task learning achieve a better performance than ImageNet-derived features for nuclear subtyping. They are also superior to those features obtained from training with SimCLR on our gland, lumen and nuclei datasets. Using MTL trained on segmentation tasks leads to stronger features as compared to single-task approaches for gland and lumen segmentation. However, using features from MTL does not surpass the performance of single task nuclear segmentation features. This may be unsurprising because the features learned for nuclear segmentation are most probably directly related to this particular task of nuclear subtyping. However, when adding patch classification as an auxiliary task during MTL, we achieve the best overall performance. MTL with auxiliary patch classification (using a ResNet18 encoder) also obtains superior features for nuclear classification as compared to the model supplied by Ciga \textit{et al.} \cite{ciga2022self}.

\begin{figure*}[t]
	\centering
\includegraphics[width=0.90\textwidth]{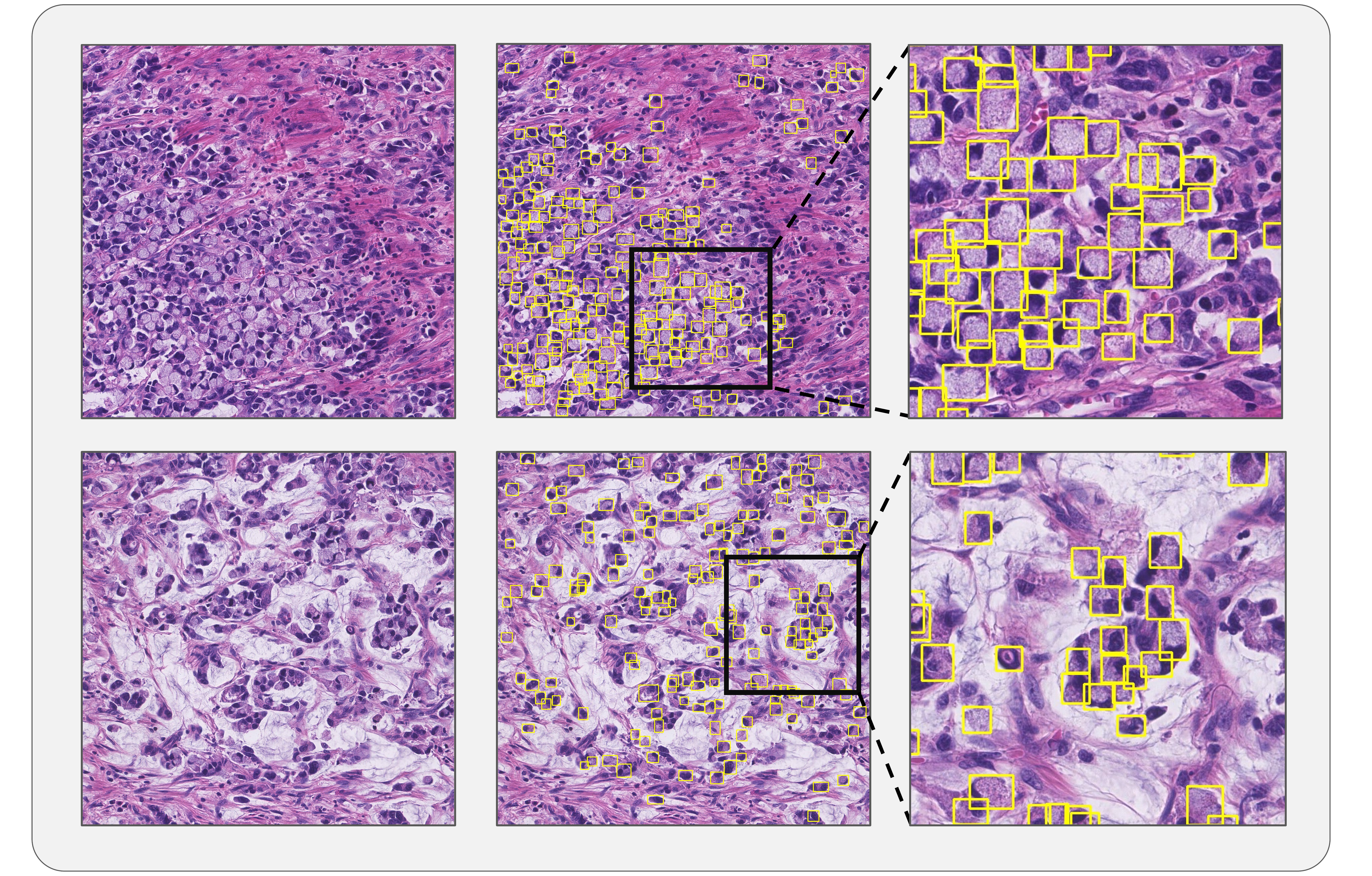}
	\caption{Signet ring cell detection visual results using the pretrained Cerberus encoder.} 
	\label{fig:signet_visual}
\end{figure*}

\begin{figure}[t]
	\centering
    \includegraphics[width=0.90\columnwidth]{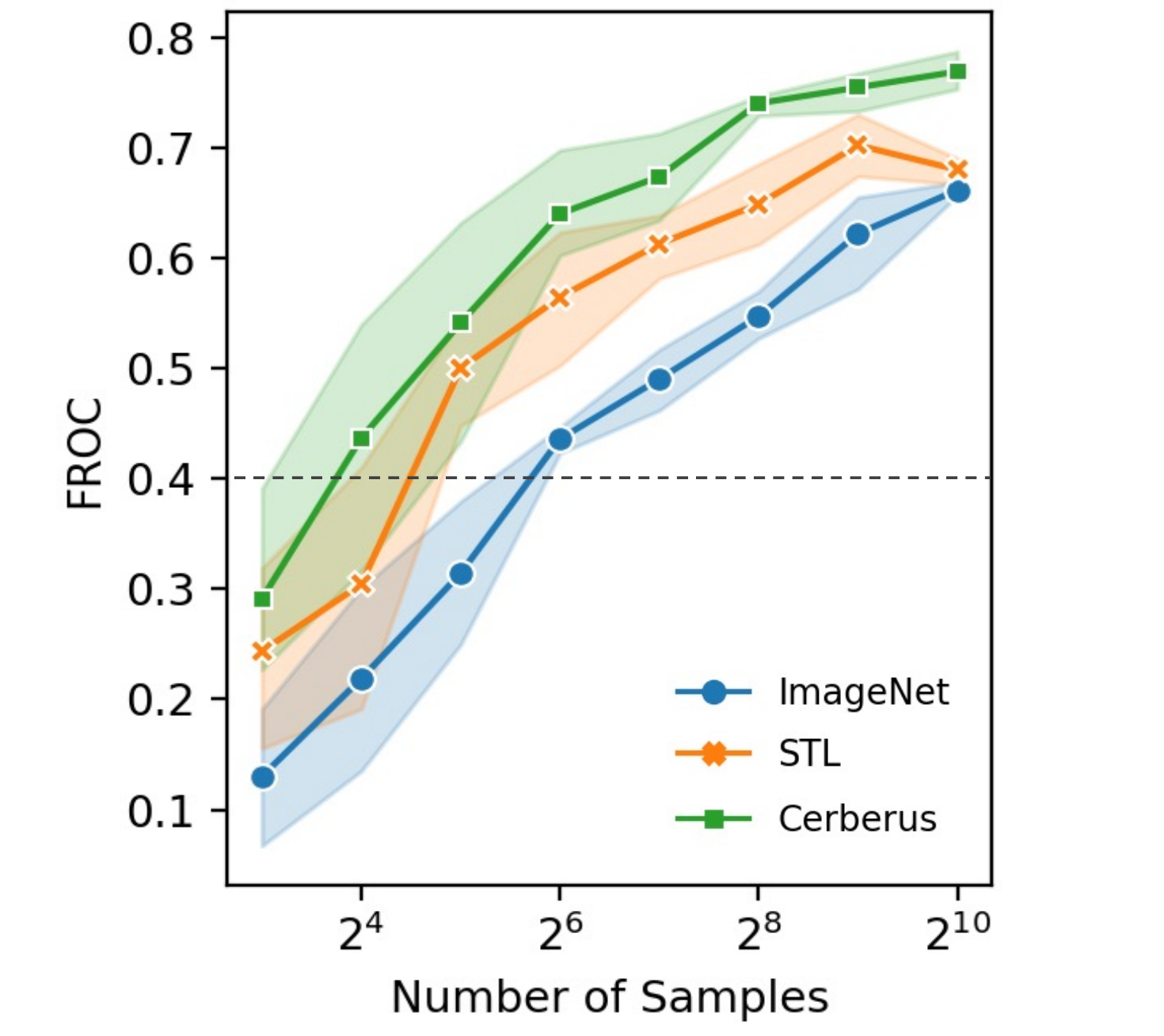}
	\caption{Signet ring cell detection performance with different numbers of input samples. We compare Cerberus features with STL (trained on nuclei) and ImageNet-based features. $FROC$: free-response receiver operating characteristic.} 
	\label{fig:signet}
\end{figure}

\subsubsection{Cerberus feature representation for object subtyping}
\label{section:mtl-nuclei-seg-results}
Cerberus is trained for binary segmentation and multi-class patch classification. However, it may be desirable to further subtype the segmentation output. For example, the output of the gland segmentation branch could be further split into benign and malignant glands, whereas the nuclear segmentation output could be split into different types of nuclei. As described in Section \ref{section:methods_subtyping}, this can be done by adding a dedicated decoder to Cerberus, which performs pixel-wise classification within objects for a particular task. Majority voting can then be used to assign a single category to each object. When utilising this strategy, the original Cerberus weights are frozen and only the additional decoder is trained. Therefore, the original model parameters remain unchanged and so performance is not compromised. Because the weights are frozen, it is essential for the encoder to have initially learned a strong feature representation to perform well on this task. In Section \ref{section:methods_subtyping}, we also describe our proposed loss function that focuses only on foreground pixels and aims to combat class imbalance.

For this experiment, we explore whether the strategy described above can be used to improve the performance achieved in Table \ref{table:feat-class-results}. The added benefit of extending Cerberus in this way is that it enables subtyping to be done jointly with other tasks and therefore a single network can be used. On the other hand, if our patch classification pipeline for nuclear classification was used, then an additional network would be needed. To give justification for our chosen subtyping network architecture and loss function, we provide an ablation study in Table \ref{table:feat-class-seg-results}. 

Rather than having an entirely separate branch for subtyping, we initially split from the binary nuclear segmentation branch after the final upsampling operation to leverage already existing features and save on computational cost. Using this strategy, we experiment with 3 different loss configurations. We first use cross-entropy (CE) and then combine CE with Dice loss to see whether this can help counter the class imbalance present in the dataset. We calculate the loss for \textit{all} pixels, which is not needed if we already perform object localisation in the segmentation branch. Therefore, we also experiment with a masked loss that computes the loss only in foreground pixels. After analysing the best loss configuration, we then see whether utilising the devoted upsampling branch can further boost the performance.

In Table \ref{table:feat-class-seg-results}, we observe that adding Dice loss to CE loss leads to an improvement in performance for all measures. There is also a further boost in the performance when using masked loss, which helps the network to learn discriminative features within the nuclei, rather than learning to separate foreground from background. However, it is evident from Table \ref{table:feat-class-results} that the performance is still inferior to the patch-classification counterpart that uses the same features. When adding the devoted upsampling branch, we observe a significant increase in performance, that even surpasses all patch-based alternatives. Therefore, it is clear that the use of a full branch for subtyping is required to ensure a strong performance.
 
 \subsubsection{Cerberus pretrained weights for object detection}
\label{section:mtl-signet-results}
To understand whether our trained multi-task model leads to better weight initialisation for new tasks, we assess whether utilising the Cerberus pretrained weights can help boost the performance of signet ring cell detection. For this, we utilise data provided as part of the DigestPath challenge, but perform additional labelling due to incomplete annotations. Here, the images with signet ring cell annotations are independent to DigestPath images used during MTL. We also want to understand whether leveraging Cerberus for enhanced weight initialisation can lead to improved data efficiency on novel tasks. For this, we train multiple signet ring cell detection models with increasing numbers of random input samples and assess the impact on performance. Specifically, we train RetinaNet \cite{lin2017focal} with a ResNet backbone to predict the bounding box around each signet ring cell. Here, the backbone is initialised with parameters either obtained from training on ImageNet, from using STL on nuclei histology data (STL-Nuclei from Table \ref{table:mtl-seg-results}) or from training Cerberus. As opposed to our nuclei subtyping experiment, we train \textit{all} model weights after initialisation. 

\begin{figure*}[t]
	\centering
    \includegraphics[width=0.92\textwidth]{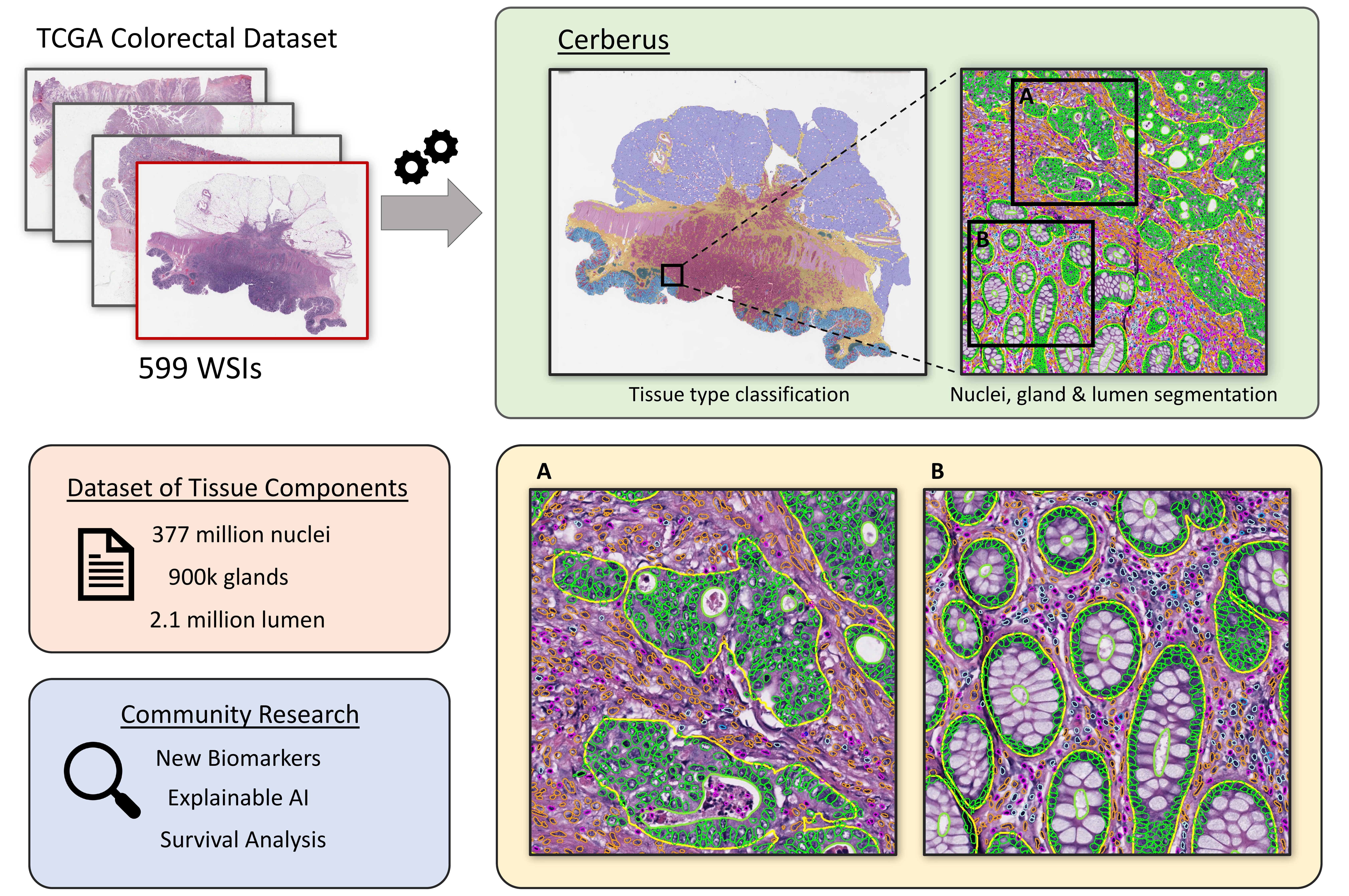}
	\caption{Overview of the pipeline for processing colorectal TCGA WSIs with Cerberus. In total, we process 599 WSIs and obtain 377 million different nuclei, 900 thousand glands and 2.1 million lumina. We release the processed dataset to enable research in the CPath community. We provide zoomed-in image regions of the corresponding boxes denoted by A and B.} 
	\label{fig:tcga}
\end{figure*}

Results from training with the different weight initialisation strategies are displayed in Figure \ref{fig:signet}, where the performance is measured in terms of area under the free-response receiver characteristic ($FROC$) curve. We choose this metric as it was used as part of the DigestPath challenge to evaluate different submissions. We repeat each experiment 4 times and therefore the plot shows the mean result along with the calculated confidence intervals. It is important to note that the $x$-axis in the plot is in log-scale. We can see that leveraging the Cerberus encoder leads to an improvement in results over ImageNet and STL pretrained weights, irrespective of the number of input samples. When we are presented with a small number of samples, then using our Cerberus encoder as a pretraining mechanism can lead to greater data efficiency. We make this clear by drawing a horizontal dashed line on the plot denoting a score of 0.4. It can be seen that much fewer samples are needed to achieve this score when using Cerberus pretrained weights. This holds potential for other tasks when not provided with a large amount of data. We display example visual results in Figure \ref{fig:signet_visual}.


 \subsubsection{Processing the TCGA colorectal database}
\label{section:tcga-results}
A major bottleneck in the development of explainable models for CPath is the localisation of various regions in the tissue. This enables the subsequent extraction of interpretable features which can then be used in downstream predictive pipelines for tasks such as biomarker discovery and survival analysis. Therefore, to accelerate research in CPath, we process all formalin-fixed paraffin-embedded colorectal WSIs from TCGA with Cerberus and make the results publicly available. We provide results for gland, lumen and nuclear segmentation, as well as patch-level tissue type classification. We also utilise our subtyping strategy (Section \ref{section:methods_subtyping}) and predict the category for each nucleus, which can enable effective modelling of the tumour micro-environment. 

In total, we localise \textbf{377 million} different types of nuclei, \textbf{900 thousand} glands and \textbf{2.1 million} lumina\footnote{This dataset is available to download at: \url{https://github.com/TissueImageAnalytics/cerberus}.}, making it by far the largest available resource of its kind in CPath. In Figure \ref{fig:tcga}, we display some example results obtained from processing sample TCGA WSIs with the proposed approach.

\subsection{Implementation and training details}
We implemented our framework with PyTorch \cite{paszke2019pytorch} version 1.9 and utilised 3 NVIDIA V100 GPUs, each with 32 GB RAM. During training, we applied the following random data augmentations: flip, rotation, Gaussian blur, median blur and colour perturbation. For segmentation tasks, networks were fed input patches of size 448$\times$448 pixels, whereas for patch classification tasks the input patch size was 144$\times$144. We trained all networks with a batch size of 9 samples per GPU (a total of 27 samples with 3 GPUs). All models were trained for 90,000 steps using Adam optimisation \cite{kingma2014adam} with an initial learning rate of 10$^{-3}$, which was reduced to 10$^{-4}$ after 70,000 steps. All networks were trained with an RGB input, normalised between 0 and 1.

\section{Discussion and conclusions}
Localisation of structures of interest in WSIs enables the extraction of interpretable features for pipelines in CPath. For this, typically a deep neural network is used per region/structure that we wish to localise. However, this may not scale well as we consider an increasing number of tissue constructs. Therefore, in this paper, we propose a multi-task learning model, named Cerberus, that performs simultaneous prediction of multiple tasks at the output of the network, without compromising on performance. In fact, we show that Cerberus achieves superior performance compared to single-task alternatives for segmentation and the learned feature representation can be leveraged to improve the performance of additional tasks via transfer learning.

Most MTL models are developed to learn a generalisable representation that is achieved by training on various independent datasets. Thereby, the collective knowledge learned can lead to more general features compared to when networks are trained on small datasets on a specific task. Despite their capability in learning a strong feature representation, datasets used in standard MTL pipelines are not necessarily aligned. For example, in CPath the datasets may have different staining, pixel resolution or belong to different tissue types. This means that, despite the network being able to make a simultaneous prediction for single image input, the outputs may not make sense. This is because each task may have its own input data assumptions. On the other hand, Cerberus ensures that each independent task-level dataset has the same input assumptions and therefore a meaningful simultaneous prediction can be made for all output branches of the network. Not only this but aligning the input helps ease network optimisation. In this paper, we use H\&E stained colon tissue at 20$\times$ objective magnification (around 0.5$\mu$m/pixel), but the method is not necessarily limited to this data configuration. Despite the fact that MTL is usually primarily used to gain a generalisable feature representation, our main focus is enabling a strong performance for simultaneous prediction by leveraging the collective knowledge from multiple available datasets. To the best of our knowledge, this is the first approach in CPath that utilises MTL with the main focus of predicting tasks simultaneously with a single network.

We have shown that Cerberus learns a strong feature representation that can be used for transfer learning. We demonstrated that the learned representation can be used to increase the performance of patch classification tasks, such as nuclear subtyping and can be used to accurately subtype objects initially segmented by the network. In addition, we show that Cerberus can be used for improved weight initialisation for tasks such as signet ring cell detection and can even lead to better data efficiency. Even though we showcased the merit of utilising Cerberus for transfer learning, there are still plenty of other applications that have not been explored. Therefore, even though Cerberus helped improve the performance of the 3 transferred tasks considered in this paper, we can't conclusively say that it \textit{always} leads to an increase in performance when applied to novel tasks. In particular, it would be interesting to see whether the feature representation learned by Cerberus can be used with other tissue types. 

For the purpose of this paper, we focussed on nuclear, gland and lumen segmentation as well as tissue type classification. Of course, it would be interesting to integrate additional tasks into our multi-task network, which can lead to a greater number of interpretable features that we can extract from the model output and may potentially increase the strength of the learned feature representation. For example, it would be interesting to consider instance segmentation of nerves and blood vessels as tasks within our framework, which would enable us to extract features relating to perineural invasion, lymphovascular invasion and tumour angiogenesis. It would also be interesting to instead use semantic segmentation of different tissue types, rather than relying on patch classification. This would enable more precise localisation of different regions, which can lead to more accurate quantification of features. As discussed previously, typical MTL pipelines in CPath do not achieve a good performance across tasks at the output of the network. However, we believe that if we ensure that the input data is aligned by the same tissue type, stain type and magnification, we can sustain a good level of performance when adding further tasks, beyond those experimented with in this paper.

During the design of our experiments, we decided which tasks to use for MTL and which to use for assessment of the learned feature representation. In theory, there is no reason why a different number of upstream and additional tasks could not be chosen for training and transfer learning of the Cerberus model. For example, we could have used signet ring cell detection during MTL and nuclear segmentation to help assess the strength of the learned features. Also, we could have added signet ring cell detection as an \textit{extra} branch during MTL. However, our chosen experimental setup allowed us to train on a large amount of data and enabled appropriate assessment of both multi-task output performance and the learned feature representation.

It would be natural to assume that due to the consideration of multiple tasks, the size of Cerberus would be very large. This is because each task has its own independent decoder. However, due to the use of relatively lightweight decoders (each containing 1.3 million parameters), the model only contains 27.1 million parameters when considering the tasks of gland segmentation, lumen segmentation, nuclear segmentation, nuclear classification and tissue type classification. This is approximately 2 times fewer than the number of parameters in MILD-Net and 3 times fewer than the number in HoVer-Net. Therefore, despite us using high-end GPUs in our experiments, this is not a requirement to fit the model into memory. When using an NVIDIA GeForce 1080 Ti GPU with a batch size of 25, the inference time for a TCGA WSI (dimensions of 28,451$\times$56,782 at a resolution of 0.5$\mu$m/pixel) was 11.7 minutes\footnote{A link to the WSI that we processed can be found here: \\ \url{https://portal.gdc.cancer.gov/files/3b88b579-a27a-4ac4-8533-48dded7d11c7}}. This will be much quicker than using separate networks for each task.

In this paper, we demonstrated how we can accurately subtype objects initially localised by our network. Specifically, we subcategorised segmented nuclei by adding an extra decoder devoted to the task of pixel classification. This can help with better quantification of important phenomena in the tumour microenvironment, which can potentially help reveal new biomarkers for downstream diagnostic and prognostic tasks. In future work, we will investigate other segmentation tasks considered within this paper. For example, our proposed subtyping approach could be used to determine whether glands are at the surface epithelium or not. We show example visual results for this in Figure \ref{fig:surf_epi}.

As part of this work, we processed 599 colorectal WSIs from the TCGA database with our proposed model. As a result, we localised a large number of nuclei and other tissue constructs, which enables the extraction of interpretable features for downstream CPath pipelines. We hope that releasing this dataset will accelerate research for the development of explainable models using interpretable features because the time-consuming localisation step is no longer required. After further inspection of the results on TCGA, we found that the model found lumen segmentation in cancerous glands particularly challenging. This may be because of the limited available context when working at 20$\times$ magnification. We chose to work at this single resolution because it was reasonable for all tasks considered by the model. However, the multi-scale analysis may yield improved performance on the task of lumen segmentation. The model sometimes also struggled to distinguish between muscle and stroma tissue for patch classification. This may be due to the very small patch size that was used, with limited context. Instead, using a semantic segmentation branch that uses a larger patch as input, as described earlier in this section, would help provide additional context and therefore improve results.

It has been described numerous times that localisation of histological primitives enables the extraction of interpretable tissue-based biomarkers, such as those shown to be predictive of clinical outcome in the literature \cite{awan2017glandular, lu2018nuclear}. However, this is only the case if the segmentation performance is good enough. For example, if we generate morphological features from poorly segmented objects, then this will lead to low-quality features, which has negative implications on generated explanations. Therefore, it is imperative to perform a large-scale assessment of the quality of the localisation across multiple slides and centres. Only then can we be confident that the features that we extract are clinically meaningful.

In future work, we would like to utilise the segmentation and classification outputs of Cerberus to extract interpretable features for explainable machine learning pipelines in CPath. For example, some features that we are able to extract from the output of Cerberus include gland, lumen and nuclear morphology; the number of inflammatory cells infiltrating into glands; intra-gland epithelial nuclei organisation and quantification of cells in the stroma. In this work, we have processed the formalin-fixed paraffin-embedded colorectal TCGA database, which has associated clinical data, including whether there exist any genetic alterations. In particular, an application of interest, especially in colon tissue, would be to understand which clinically meaningful features are associated with the microsatellite instability status. Another interesting application would be to perform survival analysis on the colorectal TCGA dataset using features, like the ones we described above.  

\section*{Acknowledgements}
SG, MJ, DS, SR, FM and NR are part of the PathLAKE digital pathology consortium, which is funded by the Data to Early Diagnosis and Precision Medicine strand of the governments Industrial Strategy Challenge Fund, managed and delivered by UK Research and Innovation (UKRI). NR was also supported by the UK Medical Research Council (grant award MR/P015476/1) and the Alan Turing Institute.

{\small
\bibliographystyle{ieee}
\bibliography{refs}

\begin{thebibliography}{10}\itemsep=-1pt

\bibitem{amgad2021nucls}
M.~Amgad, L.~A. Atteya, H.~Hussein, K.~H. Mohammed, E.~Hafiz, M.~A. Elsebaie,
  A.~M. Alhusseiny, M.~A. AlMoslemany, A.~M. Elmatboly, P.~A. Pappalardo,
  et~al.
\newblock Nucls: A scalable crowdsourcing, deep learning approach and dataset
  for nucleus classification, localization and segmentation.
\newblock {\em arXiv preprint arXiv:2102.09099}, 2021.

\bibitem{amgad2019structured}
M.~Amgad, H.~Elfandy, H.~Hussein, L.~A. Atteya, M.~A. Elsebaie, L.~S.
  Abo~Elnasr, R.~A. Sakr, H.~S. Salem, A.~F. Ismail, A.~M. Saad, et~al.
\newblock Structured crowdsourcing enables convolutional segmentation of
  histology images.
\newblock {\em Bioinformatics}, 35(18):3461--3467, 2019.

\bibitem{awan2017glandular}
R.~Awan, K.~Sirinukunwattana, D.~Epstein, S.~Jefferyes, U.~Qidwai, Z.~Aftab,
  I.~Mujeeb, D.~Snead, and N.~Rajpoot.
\newblock Glandular morphometrics for objective grading of colorectal
  adenocarcinoma histology images.
\newblock {\em Scientific reports}, 7(1):1--12, 2017.

\bibitem{bejnordi2017diagnostic}
B.~E. Bejnordi, M.~Veta, P.~J. Van~Diest, B.~Van~Ginneken, N.~Karssemeijer,
  G.~Litjens, J.~A. Van Der~Laak, M.~Hermsen, Q.~F. Manson, M.~Balkenhol,
  et~al.
\newblock Diagnostic assessment of deep learning algorithms for detection of
  lymph node metastases in women with breast cancer.
\newblock {\em Jama}, 318(22):2199--2210, 2017.

\bibitem{bulten2020automated}
W.~Bulten, H.~Pinckaers, H.~van Boven, R.~Vink, T.~de~Bel, B.~van Ginneken,
  J.~van~der Laak, C.~Hulsbergen-van~de Kaa, and G.~Litjens.
\newblock Automated deep-learning system for gleason grading of prostate cancer
  using biopsies: a diagnostic study.
\newblock {\em The Lancet Oncology}, 21(2):233--241, 2020.

\bibitem{bychkov2018deep}
D.~Bychkov, N.~Linder, R.~Turkki, S.~Nordling, P.~E. Kovanen, C.~Verrill,
  M.~Walliander, M.~Lundin, C.~Haglund, and J.~Lundin.
\newblock Deep learning based tissue analysis predicts outcome in colorectal
  cancer.
\newblock {\em Scientific reports}, 8(1):1--11, 2018.

\bibitem{caruana1997multitask}
R.~Caruana.
\newblock Multitask learning.
\newblock {\em Machine learning}, 28(1):41--75, 1997.

\bibitem{chen2017dcan}
H.~Chen, X.~Qi, L.~Yu, Q.~Dou, J.~Qin, and P.-A. Heng.
\newblock Dcan: Deep contour-aware networks for object instance segmentation
  from histology images.
\newblock {\em Medical image analysis}, 36:135--146, 2017.

\bibitem{chen2020simple}
T.~Chen, S.~Kornblith, M.~Norouzi, and G.~Hinton.
\newblock A simple framework for contrastive learning of visual
  representations.
\newblock In {\em International conference on machine learning}, pages
  1597--1607. PMLR, 2020.

\bibitem{ciga2022self}
O.~Ciga, T.~Xu, and A.~L. Martel.
\newblock Self supervised contrastive learning for digital histopathology.
\newblock {\em Machine Learning with Applications}, 7:100198, 2022.

\bibitem{francesco_ciompi_2019_3513571}
F.~Ciompi, Y.~Jiao, and J.~van~der Laak.
\newblock Lymphocyte assessment hackathon (lysto), Oct. 2019.

\bibitem{crawshaw2020multi}
M.~Crawshaw.
\newblock Multi-task learning with deep neural networks: A survey.
\newblock {\em arXiv preprint arXiv:2009.09796}, 2020.

\bibitem{da2022digestpath}
Q.~Da, X.~Huang, Z.~Li, Y.~Zuo, C.~Zhang, J.~Liu, W.~Chen, J.~Li, D.~Xu, Z.~Hu,
  et~al.
\newblock Digestpath: a benchmark dataset with challenge review for the
  pathological detection and segmentation of digestive-system.
\newblock {\em Medical Image Analysis}, page 102485, 2022.

\bibitem{diao2021human}
J.~A. Diao, J.~K. Wang, W.~F. Chui, V.~Mountain, S.~C. Gullapally,
  R.~Srinivasan, R.~N. Mitchell, B.~Glass, S.~Hoffman, S.~K. Rao, et~al.
\newblock Human-interpretable image features derived from densely mapped cancer
  pathology slides predict diverse molecular phenotypes.
\newblock {\em Nature communications}, 12(1):1--15, 2021.

\bibitem{dice1945measures}
L.~R. Dice.
\newblock Measures of the amount of ecologic association between species.
\newblock {\em Ecology}, 26(3):297--302, 1945.

\bibitem{duong2015low}
L.~Duong, T.~Cohn, S.~Bird, and P.~Cook.
\newblock Low resource dependency parsing: Cross-lingual parameter sharing in a
  neural network parser.
\newblock In {\em Proceedings of the 53rd annual meeting of the Association for
  Computational Linguistics and the 7th international joint conference on
  natural language processing (volume 2: short papers)}, pages 845--850, 2015.

\bibitem{fraz2020fabnet}
M.~Fraz, S.~A. Khurram, S.~Graham, M.~Shaban, M.~Hassan, A.~Loya, and N.~M.
  Rajpoot.
\newblock Fabnet: Feature attention-based network for simultaneous segmentation
  of microvessels and nerves in routine histology images of oral cancer.
\newblock {\em Neural Computing and Applications}, 32(14):9915--9928, 2020.

\bibitem{gamper2020pannuke}
J.~Gamper, N.~A. Koohbanani, K.~Benes, S.~Graham, M.~Jahanifar, S.~A. Khurram,
  A.~Azam, K.~Hewitt, and N.~Rajpoot.
\newblock Pannuke dataset extension, insights and baselines.
\newblock {\em arXiv preprint arXiv:2003.10778}, 2020.

\bibitem{gamper2020multi}
J.~Gamper, N.~A. Kooohbanani, and N.~Rajpoot.
\newblock Multi-task learning in histo-pathology for widely generalizable
  model.
\newblock {\em arXiv preprint arXiv:2005.08645}, 2020.

\bibitem{gamper2021multiple}
J.~Gamper and N.~Rajpoot.
\newblock Multiple instance captioning: Learning representations from
  histopathology textbooks and articles.
\newblock In {\em Proceedings of the IEEE/CVF Conference on Computer Vision and
  Pattern Recognition}, pages 16549--16559, 2021.

\bibitem{glorot2010understanding}
X.~Glorot and Y.~Bengio.
\newblock Understanding the difficulty of training deep feedforward neural
  networks.
\newblock In {\em Proceedings of the thirteenth international conference on
  artificial intelligence and statistics}, pages 249--256. JMLR Workshop and
  Conference Proceedings, 2010.

\bibitem{graham2019mild}
S.~Graham, H.~Chen, J.~Gamper, Q.~Dou, P.-A. Heng, D.~Snead, Y.~W. Tsang, and
  N.~Rajpoot.
\newblock Mild-net: Minimal information loss dilated network for gland instance
  segmentation in colon histology images.
\newblock {\em Medical image analysis}, 52:199--211, 2019.

\bibitem{graham2019rota}
S.~Graham, D.~Epstein, and N.~Rajpoot.
\newblock Rota-net: Rotation equivariant network for simultaneous gland and
  lumen segmentation in colon histology images.
\newblock In {\em European Congress on Digital Pathology}, pages 109--116.
  Springer, 2019.

\bibitem{graham2021lizard}
S.~Graham, M.~Jahanifar, A.~Azam, M.~Nimir, Y.-W. Tsang, K.~Dodd, E.~Hero,
  H.~Sahota, A.~Tank, K.~Benes, et~al.
\newblock Lizard: A large-scale dataset for colonic nuclear instance
  segmentation and classification.
\newblock In {\em Proceedings of the IEEE/CVF International Conference on
  Computer Vision}, pages 684--693, 2021.

\bibitem{graham2021conic}
S.~Graham, M.~Jahanifar, Q.~D. Vu, G.~Hadjigeorghiou, T.~Leech, D.~Snead,
  S.~E.~A. Raza, F.~Minhas, and N.~Rajpoot.
\newblock Conic: Colon nuclei identification and counting challenge 2022.
\newblock {\em arXiv preprint arXiv:2111.14485}, 2021.

\bibitem{graham2019hover}
S.~Graham, Q.~D. Vu, S.~E.~A. Raza, A.~Azam, Y.~W. Tsang, J.~T. Kwak, and
  N.~Rajpoot.
\newblock Hover-net: Simultaneous segmentation and classification of nuclei in
  multi-tissue histology images.
\newblock {\em Medical Image Analysis}, 58:101563, 2019.

\bibitem{he2015delving}
K.~He, X.~Zhang, S.~Ren, and J.~Sun.
\newblock Delving deep into rectifiers: Surpassing human-level performance on
  imagenet classification.
\newblock In {\em Proceedings of the IEEE international conference on computer
  vision}, pages 1026--1034, 2015.

\bibitem{he2016deep}
K.~He, X.~Zhang, S.~Ren, and J.~Sun.
\newblock Deep residual learning for image recognition.
\newblock In {\em Proceedings of the IEEE conference on computer vision and
  pattern recognition}, pages 770--778, 2016.

\bibitem{ioffe2015batch}
S.~Ioffe and C.~Szegedy.
\newblock Batch normalization: Accelerating deep network training by reducing
  internal covariate shift.
\newblock In {\em International conference on machine learning}, pages
  448--456. PMLR, 2015.

\bibitem{jaber2021deep}
M.~I. Jaber, B.~Song, L.~Beziaeva, C.~W. Szeto, P.~Spilman, P.~Yang, and
  P.~Soon-Shiong.
\newblock A deep learning-based iterative digital pathology annotation tool.
\newblock {\em bioRxiv}, 2021.

\bibitem{javed2020cellular}
S.~Javed, A.~Mahmood, M.~M. Fraz, N.~A. Koohbanani, K.~Benes, Y.-W. Tsang,
  K.~Hewitt, D.~Epstein, D.~Snead, and N.~Rajpoot.
\newblock Cellular community detection for tissue phenotyping in colorectal
  cancer histology images.
\newblock {\em Medical image analysis}, 63:101696, 2020.

\bibitem{kather2020pan}
J.~N. Kather, L.~R. Heij, H.~I. Grabsch, C.~Loeffler, A.~Echle, H.~S. Muti,
  J.~Krause, J.~M. Niehues, K.~A. Sommer, P.~Bankhead, et~al.
\newblock Pan-cancer image-based detection of clinically actionable genetic
  alterations.
\newblock {\em Nature Cancer}, 1(8):789--799, 2020.

\bibitem{kather2019predicting}
J.~N. Kather, J.~Krisam, P.~Charoentong, T.~Luedde, E.~Herpel, C.-A. Weis,
  T.~Gaiser, A.~Marx, N.~A. Valous, D.~Ferber, et~al.
\newblock Predicting survival from colorectal cancer histology slides using
  deep learning: A retrospective multicenter study.
\newblock {\em PLoS medicine}, 16(1):e1002730, 2019.

\bibitem{kather2016multi}
J.~N. Kather, C.-A. Weis, F.~Bianconi, S.~M. Melchers, L.~R. Schad, T.~Gaiser,
  A.~Marx, and F.~G. Z{\"o}llner.
\newblock Multi-class texture analysis in colorectal cancer histology.
\newblock {\em Scientific reports}, 6(1):1--11, 2016.

\bibitem{kingma2014adam}
D.~P. Kingma and J.~Ba.
\newblock Adam: A method for stochastic optimization.
\newblock {\em arXiv preprint arXiv:1412.6980}, 2014.

\bibitem{kirillov2019panoptic}
A.~Kirillov, K.~He, R.~Girshick, C.~Rother, and P.~Doll{\'a}r.
\newblock Panoptic segmentation.
\newblock In {\em Proceedings of the IEEE/CVF Conference on Computer Vision and
  Pattern Recognition}, pages 9404--9413, 2019.

\bibitem{kumar2019multi}
N.~Kumar, R.~Verma, D.~Anand, Y.~Zhou, O.~F. Onder, E.~Tsougenis, H.~Chen,
  P.-A. Heng, J.~Li, Z.~Hu, et~al.
\newblock A multi-organ nucleus segmentation challenge.
\newblock {\em IEEE transactions on medical imaging}, 39(5):1380--1391, 2019.

\bibitem{lin2017focal}
T.-Y. Lin, P.~Goyal, R.~Girshick, K.~He, and P.~Doll{\'a}r.
\newblock Focal loss for dense object detection.
\newblock In {\em Proceedings of the IEEE international conference on computer
  vision}, pages 2980--2988, 2017.

\bibitem{lin2019pareto}
X.~Lin, H.-L. Zhen, Z.~Li, Q.-F. Zhang, and S.~Kwong.
\newblock Pareto multi-task learning.
\newblock {\em Advances in neural information processing systems},
  32:12060--12070, 2019.

\bibitem{liu2021conflict}
B.~Liu, X.~Liu, X.~Jin, P.~Stone, and Q.~Liu.
\newblock Conflict-averse gradient descent for multi-task learning.
\newblock {\em Advances in Neural Information Processing Systems},
  34:18878--18890, 2021.

\bibitem{lu2018nuclear}
C.~Lu, D.~Romo-Bucheli, X.~Wang, A.~Janowczyk, S.~Ganesan, H.~Gilmore, D.~Rimm,
  and A.~Madabhushi.
\newblock Nuclear shape and orientation features from h\&e images predict
  survival in early-stage estrogen receptor-positive breast cancers.
\newblock {\em Laboratory investigation}, 98(11):1438--1448, 2018.

\bibitem{misra2016cross}
I.~Misra, A.~Shrivastava, A.~Gupta, and M.~Hebert.
\newblock Cross-stitch networks for multi-task learning.
\newblock In {\em Proceedings of the IEEE conference on computer vision and
  pattern recognition}, pages 3994--4003, 2016.

\bibitem{mormont2020multi}
R.~Mormont, P.~Geurts, and R.~Mar{\'e}e.
\newblock Multi-task pre-training of deep neural networks for digital
  pathology.
\newblock {\em IEEE journal of biomedical and health informatics},
  25(2):412--421, 2020.

\bibitem{naylor2018segmentation}
P.~Naylor, M.~La{\'e}, F.~Reyal, and T.~Walter.
\newblock Segmentation of nuclei in histopathology images by deep regression of
  the distance map.
\newblock {\em IEEE transactions on medical imaging}, 38(2):448--459, 2018.

\bibitem{paszke2019pytorch}
A.~Paszke, S.~Gross, F.~Massa, A.~Lerer, J.~Bradbury, G.~Chanan, T.~Killeen,
  Z.~Lin, N.~Gimelshein, L.~Antiga, et~al.
\newblock Pytorch: An imperative style, high-performance deep learning library.
\newblock {\em Advances in neural information processing systems},
  32:8026--8037, 2019.

\bibitem{ren2015faster}
S.~Ren, K.~He, R.~Girshick, and J.~Sun.
\newblock Faster r-cnn: Towards real-time object detection with region proposal
  networks.
\newblock {\em Advances in neural information processing systems}, 28, 2015.

\bibitem{ronneberger2015u}
O.~Ronneberger, P.~Fischer, and T.~Brox.
\newblock U-net: Convolutional networks for biomedical image segmentation.
\newblock In {\em International Conference on Medical image computing and
  computer-assisted intervention}, pages 234--241. Springer, 2015.

\bibitem{ruder2017overview}
S.~Ruder.
\newblock An overview of multi-task learning in deep neural networks.
\newblock {\em arXiv preprint arXiv:1706.05098}, 2017.

\bibitem{sener2018multi}
O.~Sener and V.~Koltun.
\newblock Multi-task learning as multi-objective optimization.
\newblock {\em Advances in neural information processing systems}, 31, 2018.

\bibitem{shaban2020context}
M.~Shaban, R.~Awan, M.~M. Fraz, A.~Azam, Y.-W. Tsang, D.~Snead, and N.~M.
  Rajpoot.
\newblock Context-aware convolutional neural network for grading of colorectal
  cancer histology images.
\newblock {\em IEEE transactions on medical imaging}, 39(7):2395--2405, 2020.

\bibitem{shaban2019novel}
M.~Shaban, S.~A. Khurram, M.~M. Fraz, N.~Alsubaie, I.~Masood, S.~Mushtaq,
  M.~Hassan, A.~Loya, and N.~M. Rajpoot.
\newblock A novel digital score for abundance of tumour infiltrating
  lymphocytes predicts disease free survival in oral squamous cell carcinoma.
\newblock {\em Scientific reports}, 9(1):1--13, 2019.

\bibitem{shephard2021simultaneous}
A.~J. Shephard, S.~Graham, S.~Bashir, M.~Jahanifar, H.~Mahmood, A.~Khurram, and
  N.~M. Rajpoot.
\newblock Simultaneous nuclear instance and layer segmentation in oral
  epithelial dysplasia.
\newblock In {\em Proceedings of the IEEE/CVF International Conference on
  Computer Vision}, pages 552--561, 2021.

\bibitem{sirinukunwattana2017gland}
K.~Sirinukunwattana, J.~P. Pluim, H.~Chen, X.~Qi, P.-A. Heng, Y.~B. Guo, L.~Y.
  Wang, B.~J. Matuszewski, E.~Bruni, U.~Sanchez, et~al.
\newblock Gland segmentation in colon histology images: The glas challenge
  contest.
\newblock {\em Medical image analysis}, 35:489--502, 2017.

\bibitem{srivastava2014dropout}
N.~Srivastava, G.~Hinton, A.~Krizhevsky, I.~Sutskever, and R.~Salakhutdinov.
\newblock Dropout: a simple way to prevent neural networks from overfitting.
\newblock {\em The journal of machine learning research}, 15(1):1929--1958,
  2014.

\bibitem{strezoski2019many}
G.~Strezoski, N.~v. Noord, and M.~Worring.
\newblock Many task learning with task routing.
\newblock In {\em Proceedings of the IEEE/CVF International Conference on
  Computer Vision}, pages 1375--1384, 2019.

\bibitem{tavolara2020segmentation}
T.~E. Tavolara, M.~K.~K. Niazi, G.~Beamer, and M.~N. Gurcan.
\newblock Segmentation of mycobacterium tuberculosis bacilli clusters from
  acid-fast stained lung biopsies: a deep learning approach.
\newblock In {\em Medical Imaging 2020: Digital Pathology}, volume 11320, pages
  92--98. SPIE, 2020.

\bibitem{tellez2020extending}
D.~Tellez, D.~H{\"o}ppener, C.~Verhoef, D.~Gr{\"u}nhagen, P.~Nierop,
  M.~Drozdzal, J.~Laak, and F.~Ciompi.
\newblock Extending unsupervised neural image compression with supervised
  multitask learning.
\newblock In {\em Medical Imaging with Deep Learning}, pages 770--783. PMLR,
  2020.

\bibitem{tizhoosh2018artificial}
H.~R. Tizhoosh and L.~Pantanowitz.
\newblock Artificial intelligence and digital pathology: challenges and
  opportunities.
\newblock {\em Journal of pathology informatics}, 9, 2018.

\bibitem{veeling2018rotation}
B.~S. Veeling, J.~Linmans, J.~Winkens, T.~Cohen, and M.~Welling.
\newblock Rotation equivariant cnns for digital pathology.
\newblock In {\em International Conference on Medical image computing and
  computer-assisted intervention}, pages 210--218. Springer, 2018.

\bibitem{verma2020multi}
R.~Verma, N.~Kumar, A.~Patil, N.~C. Kurian, S.~Rane, and A.~Sethi.
\newblock Multi-organ nuclei segmentation and classification challenge 2020.
\newblock {\em IEEE Transactions on Medical Imaging}, 39:1380--1391, 2020.

\bibitem{wahab2021semantic}
N.~Wahab, I.~M. Miligy, K.~Dodd, H.~Sahota, M.~Toss, W.~Lu, M.~Jahanifar,
  M.~Bilal, S.~Graham, Y.~Park, et~al.
\newblock Semantic annotation for computational pathology: Multidisciplinary
  experience and best practice recommendations.
\newblock {\em arXiv preprint arXiv:2106.13689}, 2021.

\bibitem{yang2016trace}
Y.~Yang and T.~M. Hospedales.
\newblock Trace norm regularised deep multi-task learning.
\newblock {\em arXiv preprint arXiv:1606.04038}, 2016.

\bibitem{zhang2013review}
M.-L. Zhang and Z.-H. Zhou.
\newblock A review on multi-label learning algorithms.
\newblock {\em IEEE transactions on knowledge and data engineering},
  26(8):1819--1837, 2013.

\bibitem{zhang2017survey}
Y.~Zhang and Q.~Yang.
\newblock A survey on multi-task learning.
\newblock {\em arXiv preprint arXiv:1707.08114}, 2017.

\end{thebibliography}
}

\clearpage

\appendix
\setcounter{subsection}{0}
\renewcommand{\thesubsection}{A\arabic{subsection}}
\setcounter{figure}{0}
\renewcommand{\thefigure}{A\arabic{figure}}
\setcounter{table}{0}
\renewcommand{\thetable}{A\arabic{table}}

 \section*{Appendix}
 
 \begin{table*}[h!]
\small
\begin{subtable}{1\textwidth}
\sisetup{} 
\centering
\begin{tabular}{|l|c|c|c|c|c|c|c|}
\hline
  \textbf{Features}  & \textbf{Mode} & \textbf{Epithelial} & \textbf{Lymphocyte} & \textbf{Plasma} & \textbf{Neutrophil} & \textbf{Eosinophil} & \textbf{Connective} \\ \hline
IN & Patch & 0.954 $\pm$ 0.002 & 0.742 $\pm$ 0.025 & 0.341 $\pm$ 0.033 & 0.157 $\pm$ 0.083 & 0.286 $\pm$ 0.073 & 0.728 $\pm$ 0.031 \\
SimCLR & Patch & 0.953 $\pm$ 0.004 & 0.761 $\pm$ 0.035 & 0.386 $\pm$ 0.018 & 0.146 $\pm$ 0.070 & 0.066 $\pm$ 0.033 & 0.736 $\pm$ 0.029 \\
SimCLR$^+$ & Patch & 0.971 $\pm$ 0.002 & 0.778 $\pm$ 0.029 & 0.455 $\pm$ 0.005 & 0.256 $\pm$ 0.101 & 0.302 $\pm$ 0.090 & 0.791 $\pm$ 0.020 \\
STL-Nuclei & Patch & 0.963 $\pm$ 0.001 & 0.816 $\pm$ 0.020 & 0.434 $\pm$ 0.018 & 0.242 $\pm$ 0.129 & 0.542 $\pm$ 0.111 & 0.811 $\pm$ 0.025 \\
STL-Gland & Patch & 0.958 $\pm$ 0.005 & 0.743 $\pm$ 0.027 & 0.337 $\pm$ 0.020 & 0.098 $\pm$ 0.036 & 0.393 $\pm$ 0.130 & 0.724 $\pm$ 0.049 \\
STL-Lumen & Patch & 0.948 $\pm$ 0.005 & 0.726 $\pm$ 0.044 & 0.330 $\pm$ 0.024 & 0.103 $\pm$ 0.046 & 0.308 $\pm$ 0.117 & 0.703 $\pm$ 0.036 \\
MTL & Patch & 0.968 $\pm$ 0.001 & 0.793 $\pm$ 0.019 & 0.410 $\pm$ 0.019 & 0.156 $\pm$ 0.057 & 0.399 $\pm$ 0.137 & 0.793 $\pm$ 0.031 \\
IN-MTL & Patch & 0.969 $\pm$ 0.003 & 0.789 $\pm$ 0.023 & 0.427 $\pm$ 0.027 & 0.208 $\pm$ 0.099 & 0.482 $\pm$ 0.133 & 0.768 $\pm$ 0.050 \\
IN-MTL+PC & Patch & 0.974 $\pm$ 0.014 & 0.807 $\pm$ 0.049 & \textbf{0.463 $\pm$ 0.079} & 0.228 $\pm$ 0.117 & 0.608 $\pm$ 0.097 & 0.828 $\pm$ 0.058 \\ \hdashline
IN-MTL+PC$^*$ & Patch & 0.970 $\pm$ 0.002 & \textbf{0.812 $\pm$ 0.029} & 0.439 $\pm$ 0.021 & 0.233 $\pm$ 0.089 & 0.513 $\pm$ 0.129 & 0.798 $\pm$ 0.034 \\
Ciga \textit{et al.} \cite{ciga2022self} & Patch & 0.962 $\pm$ 0.004 & 0.763 $\pm$ 0.017 & 0.403 $\pm$ 0.011 & 0.256 $\pm$ 0.078 & 0.553 $\pm$ 0.055 & 0.755 $\pm$ 0.035 \\ \hdashline 
Cerberus & Pixel & \textbf{0.984 $\pm$ 0.002} & 0.775 $\pm$ 0.019 & 0.430 $\pm$ 0.017 & \textbf{0.358 $\pm$ 0.134} & \textbf{0.635 $\pm$ 0.065} & \textbf{0.886 $\pm$ 0.014}  \\ \hline 
\end{tabular}
\caption{Cross validation results.}
\end{subtable}

\bigskip
\begin{subtable}{1\textwidth}
\sisetup{} 
\centering
\begin{tabular}{|l|c|c|c|c|c|c|c|}
\hline
  \textbf{Features}  & \textbf{Mode} & \textbf{Epithelial} & \textbf{Lymphocyte} & \textbf{Plasma} & \textbf{Neutrophil} & \textbf{Eosinophil} & \textbf{Connective} \\ \hline
IN & Patch & 0.682 $\pm$ 0.017 & 0.517 $\pm$ 0.026 & 0.275 $\pm$ 0.028 & 0.055 $\pm$ 0.019 & 0.334 $\pm$ 0.047 & 0.483 $\pm$ 0.053 \\
SimCLR & Patch & 0.885 $\pm$ 0.022 & 0.674 $\pm$ 0.026 & 0.401 $\pm$ 0.049 & 0.029 $\pm$ 0.011 & 0.044 $\pm$ 0.009 & 0.560 $\pm$ 0.096 \\
SimCLR$^+$ & Patch & 0.917 $\pm$ 0.022 & 0.644 $\pm$ 0.036 & 0.336 $\pm$ 0.057 & 0.049 $\pm$ 0.023 & 0.351 $\pm$ 0.066 & 0.630 $\pm$ 0.030 \\
STL-Nuclei & Patch & 0.932 $\pm$ 0.009 & 0.743 $\pm$ 0.043 & 0.395 $\pm$ 0.045 & 0.060 $\pm$ 0.038 & 0.571 $\pm$ 0.032 & 0.729 $\pm$ 0.048 \\
STL-Gland & Patch & 0.912 $\pm$ 0.018 & 0.606 $\pm$ 0.132 & 0.367 $\pm$ 0.035 & 0.016 $\pm$ 0.006 & 0.450 $\pm$ 0.051 & 0.620 $\pm$ 0.099 \\
STL-Lumen & Patch & 0.900 $\pm$ 0.016 & 0.681 $\pm$ 0.065 & 0.326 $\pm$ 0.056 & 0.017 $\pm$ 0.005 & 0.211 $\pm$ 0.139 & 0.575 $\pm$ 0.095 \\
MTL & Patch & 0.939 $\pm$ 0.011 & 0.745 $\pm$ 0.010 & 0.429 $\pm$ 0.018 & 0.034 $\pm$ 0.010 & 0.496 $\pm$ 0.044 & 0.708 $\pm$ 0.043 \\
IN-MTL & Patch & 0.933 $\pm$ 0.016 & 0.657 $\pm$ 0.125 & 0.432 $\pm$ 0.037 & 0.040 $\pm$ 0.017 & 0.480 $\pm$ 0.037 & 0.698 $\pm$ 0.038 \\
IN-MTL+PC & Patch & 0.937 $\pm$ 0.018 & 0.714 $\pm$ 0.047 & 0.405 $\pm$ 0.018 & 0.062 $\pm$ 0.010 & 0.493 $\pm$ 0.030 & 0.726 $\pm$ 0.051 \\ \hdashline
IN-MTL+PC$^*$ & Patch & 0.939 $\pm$ 0.014 & \textbf{0.750 $\pm$ 0.028} & \textbf{0.510 $\pm$ 0.027} & 0.048 $\pm$ 0.015 & 0.559 $\pm$ 0.029 & 0.710 $\pm$ 0.079 \\
Ciga \textit{et al.} \cite{ciga2022self} & Patch & 0.931 $\pm$ 0.008 & 0.713 $\pm$ 0.020 & 0.329 $\pm$ 0.042 & 0.030 $\pm$ 0.006 & 0.558 $\pm$ 0.061 & 0.682 $\pm$ 0.033 \\ \hdashline
Cerberus & Pixel & \textbf{0.974 $\pm$ 0.004} & 0.688 $\pm$ 0.036 & 0.447 $\pm$ 0.029 & \textbf{0.116 $\pm$ 0.033} & \textbf{0.720 $\pm$ 0.006} & \textbf{0.858 $\pm$ 0.014} \\ \hline 
\end{tabular}
\caption{External test results.}
\end{subtable}
\caption{Breakdown of nuclear classification performance (average precision) for each category. Ciga \textit{et al.} \cite{ciga2022self} uses SimCLR for model training. The difference between this and SimCLR/SimCLR$^+$ reported on rows 2 and 3 is the data used during optimisation. Here, SimCLR uses the same data used during MTL and SimCLR$^+$ uses over 2 million colon H\&E image patches from many datasets.}
\label{table:feat-class-breakdown}
\end{table*}

\subsection*{Breakdown of nuclear subtyping results}
To provide a more detailed analysis, we give the $AP$ per nuclear category for subtyping tasks in Table \ref{table:feat-class-breakdown}. Here, we show the results for patch-based and pixel-based subtyping, as described in Sections \ref{section:mtl-nuclei-results} and \ref{section:mtl-nuclei-seg-results} respectively. For patch classification, we provide all methods that were given in the main paper, whereas for pixel-level classification we only provide the results obtained by Cerberus. 

When using our multi-task features with auxiliary patch classification of the tissue type, we regularly outperform all other methods for patch-based subtyping. This performance is further improved when using pixel-level subtyping. We observe that, in particular, using our pixel-based subtyping method had the greatest impact on the neutrophil and eosinophil performance. For these classes, there are much fewer examples and therefore our proposed loss function may have been able to better deal with the class imbalance in the dataset. Nevertheless, the performance of epithelial cell subtyping also improved with pixel-level subtyping. Despite this, the neutrophil performance is still relatively low on the external test set. In future work, we would like to explore why this is low and understand how we can further improve the performance.

\subsection*{Subtyping gland segmentation output}
The subtyping method that we described in Section \ref{section:methods_subtyping} is not limited to nuclear classification. We use the same method to classify glands as either surface epithelium or glands located within the tissue. In fact, the surface epithelium is not typically denoted as being glandular and therefore subtyping in this way is important. Also, analysis of glandular morphology should typically be performed at glands that have been cut in the transverse plane and not at the surface epithelium. Therefore, this subcategorisation of the gland segmentation output also enables the extraction of better features for subsequent analysis.

\begin{figure*}[t]
	\centering
    \includegraphics[width=0.80\textwidth]{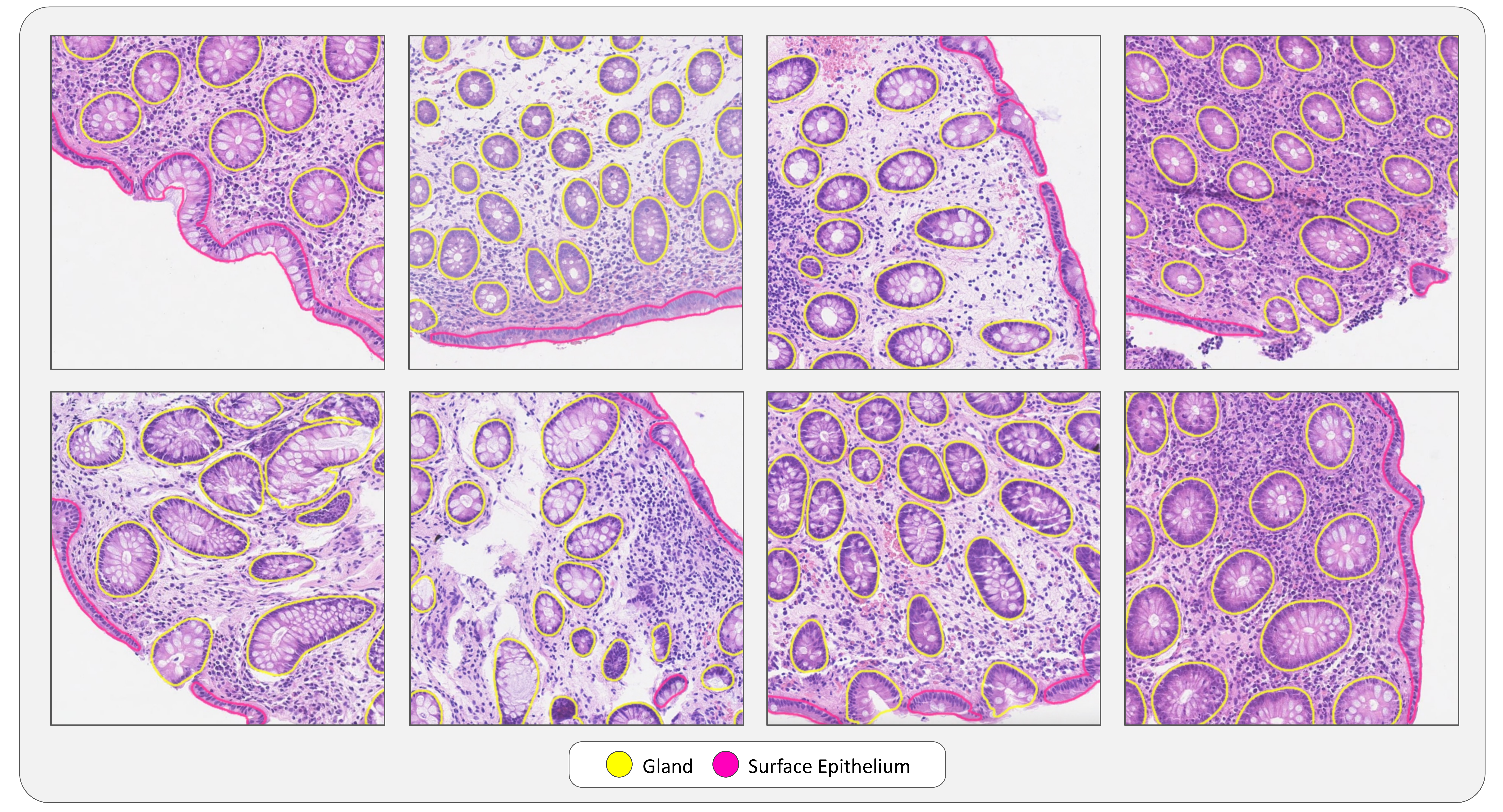}
	\caption{Visual results demonstrating classification of glands as either surface epithelium or not.} 
	\label{fig:surf_epi}
\end{figure*}

In Figure \ref{fig:surf_epi} we display some example visual results, where we observe that our method can well differentiate surface epithelium from glands within the tissue.

\newpage

\end{document}